\documentclass[aps,
twocolumn,pre,superscriptaddress
]{revtex4}
\usepackage{amsmath}
\usepackage{amssymb}
\usepackage{graphicx}
\usepackage{color}
\usepackage{enumitem}

\begin{document}

\title{Solvable macroscopic models for populations of type~I excitability neurons with Gaussian noise}
\date{\today}
\author{Maria V.\ Ageeva}
\email{ageeva_mv00@mail.ru}
\affiliation{Institute of Continuous Media Mechanics, Ural Branch of RAS, Acad.\ Korolev Street 1,
 614013 Perm, Russia}
\affiliation{Institute of Physics and Mathematics, Perm State University, Bukirev Street 15,
 614990 Perm, Russia}

\author{Denis S.\ Goldobin}
\email{Denis.Goldobin@gmail.com}
\affiliation{Institute of Continuous Media Mechanics, Ural Branch of RAS, Acad.\ Korolev Street 1,
 614013 Perm, Russia}
\affiliation{Institute of Physics and Mathematics, Perm State University, Bukirev Street 15,
 614990 Perm, Russia}
\affiliation{Department of Control Theory, Lobachevsky State University of Nizhny Novgorod, Gagarin Avenue 23,
 603022 Nizhny Novgorod, Russia}

\begin{abstract}
The dynamics of type I excitability neurons near the bifurcation point are governed by the model of quadratic integrate-and-fire neuron (QIF). The model of QIF subject to white Gaussian noise is abundant in statistical physics, since it not only reasonably reproduces the dynamics of some biological systems but also emerges in the problems of stochastic stability and in quantum mechanics of one-dimensional systems with quenched disorder.
We derive the solution for the mean interspike interval of a QIF subject to additive Gaussian noise in terms of analytic functions, from the first passage time problem. On the basis of this solution, we construct analytical expressions for the firing rate and population-mean membrane potential in terms of the real and imaginary parts of a certain universal analytic function. With this function, we generalize theoretical results to heterogeneous populations with Lorentzian-distributed parameters. Implementation of theoretical results is demonstrated and examined for two problems: the interplay between chemical and electrical synapses, and the emergence of effective endogenous noise in sparse inhibitory networks. For the latter case, it is possible to cover the scenarios where the effective noise intensity depends on Lorentzian-distributed parameters, which is not feasible directly with the diffusion approximation.
\end{abstract}

\maketitle

\section{Introduction}
For type~I excitability neurons, the normal form of dynamics near the bifurcation point of transition between the regime of periodic spiking and the silent (excitable) state is given by the model of quadratic integrate-and-fire neuron (QIF)~\cite{Ermentrout-Kopell-1986,Izhikevich-2007}.
This model reasonably reproduces the dynamics of some biological systems~\cite{Ermentrout-Kopell-1986,Izhikevich-2007}.
The problem of calculating the Lyapunov exponent~\cite{Pikovsky-Politi-LE_book-2016} for the temporal dynamics of noisy systems~\cite{Klyatskin-2017,Klyatskin-1975,Zillmer-Pikovsky-2003,Permyakova-Goldobin-2023,Khasminskii-2012,Goldobin-etal-2017} and for spatial patterns in one-dimensional systems with quenched disorder~\cite{Frisch-Lloyd-1960,Halperin-1965,Lifshitz-Gredeskul-Pastur-1988, Zillmer-Pikovsky-2005,Derrida-Hilhorst-1983,Mello-Robledo-1993, Goldobin-Shklyaeva-2013,Goldobin-Shklyaeva-2009,Goldobin-2010} is addressed using equivalent or similar mathematical models, albeit with different interpretations of variables and parameters in some cases.
Typically, for these problems, one can adopt the diffusion approximation (DA) well-established in both mathematical neuroscience~\cite{Capocelli-Ricciardi-1971} and quantum mechanics~\cite{Frisch-Lloyd-1960}. Within this approximation, the noise is represented by continuous $\delta$-correlated Gaussian noise.

For QIFs, the primary quantifiers of mean-field activity in a population of coupled units are the firing rate and the population-mean membrane potential. Until recently~\cite{Ageeva-Goldobin-PREL-2026}, the literature~\cite{Hirsch-Huberman-Scalapino-1982,Lindner-Longtin-Bulsara-2003,Lifshitz-Gredeskul-Pastur-1988,Pikovsky-Politi-LE_book-2016} reported only quadrature forms of solutions for these quantifiers. In this paper, we present an analytical solution which offers significant advantages~\cite{Ageeva-Goldobin-PREL-2026} compared to the quadrature form of solution~\cite{Lindner-Longtin-Bulsara-2003} and the partial analytical solution presented in~\cite{Volo-etal-2022}.

The paper is organized as follows. In Sec.\,\ref{sec2:QIF}, we introduce the QIF model and its Fokker--Planck equation. For a homogeneous population of QIFs, we derive the mean interspike interval as an analytical solution to the first passage time problem in Sec.\,\ref{ssec31:FPT}, from which we calculate the firing rate and population-mean membrane potential in Secs.\,\ref{ssec32:FRate} and \ref{ssec33:MV}. In Sec.\,\ref{ssec34:MPR}, we recast these results in a form natural for the order parameter introduced by Montbri\'o, Paz\'o, and Roxin~\cite{Montbrio-Pazo-Roxin-2015}, also known as the first pseudocumulant~\cite{Goldobin-Volo-Torcini-2021}.
In Sec.\,\ref{sec4:Heter}, this form allows us to generalize the analytical result to heterogenous populations with Lorentzian-distributed parameters.
In Sec.\,\ref{sec5:mean-field}, we apply the analytical result to two examples: networks with chemical and electrical synapses and sparse inhibitory networks.
In Sec.\,\ref{sec6:Concl}, the conclusions are drawn.
In Appendixes~\ref{secA1} and \ref{secA2}, we demonstrate how the Residue theorem can be applied to the dif\-fu\-sion-approximation models, where the direct implementation with heterogeneous effective noise intensity is inadmissible, and explain the mathematical connections between neuroscience and quantum mechanics problems.

\section{Quadratic integrate-and-fire neurons with Gaussian noise}\label{sec2:QIF}
We consider a population of QIFs~\cite{Ermentrout-Kopell-1986,Izhikevich-2007}
with independent individual noises:
\begin{align}
&\dot{V}_j=V_j^2+I_j(t)\,,
\label{eq101}
\\
&I_j=A_j+\sqrt{D}\zeta_j(t)\,,
\label{eq102}
\end{align}
where $V_j$ represents the membrane potential of the $j$th neuron, $I_j$ is the synaptic input current, $A_j$ is the DC synaptic input current; $\zeta_j(t)$ are independent normalized Gaussian noise signals: $\langle\zeta_j(t)\,\zeta_m(t+t^\prime)\rangle=2\delta_{jm}\delta(t^\prime)$.
When the potential $V_j$ reaches a threshold value $B_+$, it is reset to $-B_-$ and a synaptic pulse is emitted. For $B_\pm\to\infty$, the results become insensitive to the specific values of $B_\pm$, allowing us to adopt the limiting case $B_\pm\to\infty$ for simplicity.

The probability density function (PDF) $P_A(V,t)$ for the subpopulation with identical values of $A_j=A$ obeys the Fokker--Planck equation:
\begin{equation}
\frac{\partial P_A(V,t)}{\partial t} +\frac{\partial}{\partial V}\Big[(A+V^2)P_A(V,t)\Big]=D\frac{\partial^2P_A(V,t)}{\partial V^2}.
\label{eq103}
\end{equation}
In the following sections, for the time-independent states of Eq.\,(\ref{eq103}), we derive the main result of this paper: the firing rate $r$ and mean potential $v$, which constitute the Montbri\'o--Paz\'o--Roxin (MPR) order parameter $W=\pi r-iv$~\cite{Montbrio-Pazo-Roxin-2015}, also known as the first pseudocumulant~\cite{Goldobin-Volo-Torcini-2021}, can be calculated~\cite{Ageeva-Goldobin-PREL-2026} as
\begin{align}
W&=\pi r-iv=D^{1/3}\mathcal{W}(\mathcal{A})\,,
\label{eqW1}
\\
\mathcal{W}(\mathcal{A})&=\frac{\sqrt{\mathcal{A}}}{i}
\frac{\mathrm{J}_{2/3}(e^{i\pi}\chi)-\mathrm{J}_{-2/3}(e^{i\pi}\chi)}
{\mathrm{J}_{1/3}(e^{i\pi}\chi)+\mathrm{J}_{-1/3}(e^{i\pi}\chi)}\,,
\label{eqW2}
\\
\mathcal{A}&=
\frac{A}{D^{2/3}}\,,\qquad
\chi=\frac23\mathcal{A}^{3/2}.
\end{align}
This result can be extended to cases involving Lorentzian parameter distributions [see Eq.\,(\ref{eq303})] and can be applied to solve self-consistent problems related to the mean-field states of populations of coupled QIFs.

\section{Homogeneous population}\label{sec3:Homo}
\subsection{First passage time problem for the interspike interval statistics}
\label{ssec31:FPT}
The statistical characterization of the interspike intervals is delivered by the first passage time problem (FPTP) \cite{Lindner-Longtin-Bulsara-2003,Capocelli-Ricciardi-1971,Gardiner-1997,Goldobin-2024}.
The solution to the FPTP comes from the Kolmogorov equation, also known as the backward Fokker--Planck equation~\footnote{See Eqs.\,(5.2.154) and (5.2.156) in~\cite{Gardiner-1997}, where the different normalization of noise is adopted and requires to substitute ``$B(x)/2$'' with $D$, or a detailed derivation for a symmetric $\alpha$-stable noise in~\cite{Goldobin-2024}, where one can take $\alpha=2$ for a Gaussian noise.}; for stochastic system~(\ref{eq101}), we obtain
\begin{equation}
(A+V_0^2)\frac{\partial\langle{T^m(V_0)}\rangle}{\partial V_0}
+D\frac{\partial^2\langle{T^m(V_0)}\rangle}{\partial V_0^2}
=-m\langle{T^{m-1}(V_0)}\rangle\,,
\label{eq501}
\end{equation}
where $m=1,2,3,\dots$ and $T(V_0)$ is the time of the first passage through $V=+\infty$ for a QIF starting from $V_0$. Eq.\,(\ref{eq501}) is complemented with the boundary conditions
\begin{equation}
\left.\frac{\partial\langle{T^m(V_0)}\rangle}{\partial V_0}\right|_{V_0=-\infty}\!\!=0\,,
\quad
\langle{T^m(V_0=+\infty)}\rangle=0\,.
\label{eq502}
\end{equation}
We obtain
\begin{equation}
\langle{T^m(V_0)}\rangle=D^{-m/3}\langle{\mathcal{T}^m(\mathcal{V}_0)}\rangle\,,
\label{eq503}
\end{equation}
where dimensionless $\mathcal{T}=D^{1/3}T$ and $\mathcal{V}_0=V_0/D^{1/3}$.
The interspike interval corresponds to the travel from $\mathcal{V}_0=-\infty$;
\[
\langle{\mathcal{T}^m(\mathcal{V}_0=-\infty)}\rangle=:\langle\tau^m\rangle\,.
\]

Now we adopt the approach originally introduced by Halperin in quantum mechanics, for the one-dimensional Anderson localization problem~\cite{Halperin-1965}, to solve the FPTP in dimensionless terms of $\mathcal{A}$, $\mathcal{V}_0$, and $\mathcal{T}$. For $f(\mathcal{V}_0)=-\frac{\partial}{\partial\mathcal{V}_0}\langle{\mathcal{T}(\mathcal{V}_0)}\rangle$, we can write
\begin{equation}
\langle\tau\rangle=\int\nolimits_{-\infty}^{+\infty}f(\mathcal{V}_0)\,\mathrm{d}\mathcal{V}_0\,,
\label{eq504}
\end{equation}
and Eq.\,(\ref{eq501}) and the boundary conditions~(\ref{eq502}) at infinity read
\begin{equation}
(\mathcal{A}+\mathcal{V}_0^2)f(\mathcal{V}_0)+\frac{\mathrm{d}}{\mathrm{d}\mathcal{V}_0}f(\mathcal{V}_0)=1\,,
\quad
f(\pm\infty)=0\,.
\label{eq505}
\end{equation}
In Fourier space, with
 $F(q)=\int_{-\infty}^{+\infty}f(\mathcal{V}_0)\,e^{iq\mathcal{V}_0}\mathrm{d}\mathcal{V}_0$ and
 $f(\mathcal{V}_0)=(2\pi)^{-1}\int_{-\infty}^{+\infty}F(q)\,e^{-iq\mathcal{V}_0}\mathrm{d}q$,
problem~(\ref{eq505}) transforms into
\begin{equation}
\left(\mathcal{A}-\frac{\mathrm{d}^2}{\mathrm{d}q^2}\right)F(q) -iqF(q)=2\pi\delta(q)\,,
\label{eq506}
\end{equation}
with properties:
\begin{equation}
\langle\tau\rangle=F(q=0),
\label{eq:tauF0}
\end{equation}
$F(\pm\infty)=0$, and, as $f(\mathcal{V}_0)$ is real, $F(-q)=F^\ast(q)$.

By integrating Eq.\,(\ref{eq506}) over the vicinity of the origin, we obtain  $-\frac{\mathrm{d}F(q)}{\mathrm{d}q}\big|_{-0}^{+0}=2\pi$; with the symmetry property $F(-q)=F^\ast(q)$, this result can be rewritten as
\begin{equation}
\left.\mathrm{Re}\frac{\mathrm{d}F(q)}{\mathrm{d}q}\right|_{q=+0}=-\pi\,.
\label{eq507}
\end{equation}
For $q>0$, we introduce $s=iq-\mathcal{A}$ and, according to Eq.\,(\ref{eq506}), the function $F(q>0)=G(s)=G(iq-\mathcal{A})$, where $G^{\prime\prime}(s)=sG(s)$ and $G(s)\to 0$ as $s\to +i\infty$. The general solution to the last equation is given by Airy functions~\cite{Abramowitz}: $G(s)=a\mathrm{Ai}(s)+b\mathrm{Bi}(s)$.
For $|s_1|\to\infty$, Airy functions have the asymptotic form~\cite{Abramowitz}
\begin{align}
\mathrm{Ai}(-s_1)&\approx\frac{1}{\sqrt{\pi}s_1^{1/4}}\sin\left(\frac{\pi}{4}+\frac{2}{3}s_1^{3/2}\right)\,,
\nonumber\\
\mathrm{Bi}(-s_1)&\approx\frac{1}{\sqrt{\pi}s_1^{1/4}}\cos\left(\frac{\pi}{4}+\frac{2}{3}s_1^{3/2}\right)
\nonumber
\end{align}
in the sector $-2\pi/3<\arg{s_1}<2\pi/3$; and the argument $iq-\mathcal{A}=e^{i\pi}s_1$ for $q\to+\infty$ falls into this sector. Hence,
\begin{align}
\mathrm{Ai}(iq-\mathcal{A})&
\approx
\frac{e^{-i\frac{\pi}{8}}}{2\sqrt{\pi}q^{1/4}}e^{\frac{1-i}{3}\sqrt{2}q^{3/2} +\frac{1+i}{\sqrt{2}}\mathcal{A}\sqrt{q} 
}
,
\nonumber\\
\mathrm{Bi}(iq-\mathcal{A})&
\approx
\frac{i e^{-i\frac{\pi}{8}}}{2\sqrt{\pi}q^{1/4}}e^{\frac{1-i}{3}\sqrt{2}q^{3/2} +\frac{1+i}{\sqrt{2}}\mathcal{A}\sqrt{q} 
}
.
\nonumber
\end{align}
The condition $\lim_{q\to+\infty}F(q)=0$ requires $a+ib=0$ to cancel the leading exponentially growing terms of $\mathrm{Ai}(iq-\mathcal{A})$ and $\mathrm{Bi}(iq-\mathcal{A})$ and prevent divergence of $F(q)$. Therefore, $F(q)=a[\mathrm{Ai}(iq-\mathcal{A})+i\mathrm{Bi}(iq-\mathcal{A})]$. Relation~(\ref{eq:tauF0}) requires $\mathrm{Im}\,F(0)=0$. With decomposition $a=a_r+ia_i$, this equation and condition~(\ref{eq507}) yield equation system for $a_r$ and $a_i$:
\begin{align}
a_i\mathrm{Ai}(-\mathcal{A})+a_r\mathrm{Bi}(-\mathcal{A})&=0\,,
\label{eq508}
\\
a_i\mathrm{Ai}^\prime(-\mathcal{A})+a_r\mathrm{Bi}^\prime(-\mathcal{A})&=\pi\,.
\label{eq509}
\end{align}
Given that the Wronskian $\mathrm{Ai}^\prime(s)\mathrm{Bi}(s) -\mathrm{Bi}^\prime(s)\mathrm{Ai}(s)=-1/\pi$ \cite{Abramowitz}, the solution to the system of equations~(\ref{eq508}) and (\ref{eq509}) reads $a_r=\pi^2\mathrm{Ai}(-\mathcal{A})$ and $a_i=-\pi^2\mathrm{Bi}(-\mathcal{A})$. Finally, the solution to the FPTP in Fourier space reads
\begin{equation}
F(q)=\pi^2[\mathrm{Ai}(-\mathcal{A})-i\mathrm{Bi}(-\mathcal{A})] [\mathrm{Ai}(iq-\mathcal{A})+i\mathrm{Bi}(iq-\mathcal{A})]\,;
\label{eq510}
\end{equation}
and the dimensionless mean interspike interval is
\begin{equation}
\langle\tau\rangle=F(0)=\pi^2\left\{[\mathrm{Ai}(-\mathcal{A})]^2+[\mathrm{Bi}(-\mathcal{A})]^2\right\}\,.
\label{eq511}
\end{equation}

By expressing the Airy functions in terms of the Bessel functions of order $1/3$~\cite{Abramowitz}, we can rewrite Eq.\,(\ref{eq511}) as
\begin{widetext}
\begin{align}
&\langle\tau\rangle=
\left\{
\begin{array}{cc}
\displaystyle
\frac{4\pi^2(-\mathcal{A})}{9}\left[
 \mathrm{I}_{1/3}^2(\chi_-) +\mathrm{I}_{-1/3}^2(\chi_-) +\mathrm{I}_{1/3}(\chi_-)\,\mathrm{I}_{-1/3}(\chi_-)\right] & \mbox{ for } \mathcal{A}<0\,, \\[10pt]
\displaystyle
\frac{4\pi^2\mathcal{A}}{9}\!\left[\mathrm{J}_{1/3}^2(\chi) +\mathrm{J}_{-1/3}^2(\chi) -\mathrm{J}_{1/3}(\chi)\,\mathrm{J}_{-1/3}(\chi)\right] & \mbox{ for } \mathcal{A}\ge0\,,
\end{array}
\right.
\label{eq512}
\end{align}
\end{widetext}
where $\chi=(2/3)\mathcal{A}^{3/2}$ and $\chi_-=(2/3)(-\mathcal{A})^{3/2}$, $\mathrm{J}_n$ and $\mathrm{I}_n$ are the $n$th order Bessel function of the first kind and the modified one, respectively.
It is important to note that expressions (\ref{eq512}) for positive and negative $\mathcal{A}$ are analytic continuations of each other, that is they are identical analytic functions. However, we present explicit forms with only nonnegative real argument values to avoid incorrect choices of branches for the multivalued (multiple-sheet) fractional-order Bessel functions on the complex plane.

\subsection{Firing rate for a time-independent macroscopic state}
\label{ssec32:FRate}
Due to ergodicity, the firing rate can be expressed via $\langle\tau\rangle$ ({\it e.g.}, \cite{Goldobin-2024}): the average over a large number $M$ of spikes is
\begin{align}
  r & =\frac{M}{T_1(-\infty)+T_2(-\infty)+\cdots+T_M(-\infty)}
\nonumber\\
   & = \frac{D^{1/3}}{\langle\tau\rangle}\,,
\label{eq:r_tau}
\end{align}
where we used scaling (\ref{eq503}). Substituting Eq.\,(\ref{eq512}) into (\ref{eq:r_tau}), we can introduce dimensionless $\mathcal{R}(\mathcal{A})$\,:
\begin{align}
&\frac{r}{D^{1/3}}\equiv\mathcal{R}\left({\frac{A}{D^{2/3}}}\right)
\nonumber\\
&
=\left\{
\begin{array}{cc}
\frac{\frac{9}{4\pi^2(-\mathcal{A})}}{
 \mathrm{I}_{1/3}^2(\chi_-) +\mathrm{I}_{-1/3}^2(\chi_-) +\mathrm{I}_{1/3}(\chi_-)\,\mathrm{I}_{-1/3}(\chi_-)} &\mbox{for } \mathcal{A}<0\,, \\
\frac{\frac{9}{4\pi^2\mathcal{A}}}{ \mathrm{J}_{1/3}^2(\chi) +\mathrm{J}_{-1/3}^2(\chi) -\mathrm{J}_{1/3}(\chi)\,\mathrm{J}_{-1/3}(\chi)} &\mbox{for } \mathcal{A}\ge0\,.
\end{array}
\right.
\label{eq204}
\end{align}
Again, the expressions for positive and negative $\mathcal{A}$ are analytic continuations of one another, but we present explicit forms with only nonnegative real argument values to avoid incorrect choices of branches for the multivalued Bessel functions.

Using the properties of Bessel functions, $\mathrm{I}_\alpha(Z)=i^{-\alpha}\mathrm{J}_\alpha(iZ)$ and $\mathrm{J}_\alpha(e^{in\pi}Z)=e^{i\alpha n\pi}\mathrm{J}_\alpha(Z)$~\cite{Abramowitz}, 
we calculate for the denominator
\begin{align}
&\big[\mathrm{J}_{1/3}(\chi)\big]^2 +\big[\mathrm{J}_{-1/3}(\chi)\big]^2 -\mathrm{J}_{1/3}(\chi)\,\mathrm{J}_{-1/3}(\chi)
\nonumber\\[5pt]
&=\mathrm{J}_{1/3}(e^{i\pi}\chi)\,\mathrm{J}_{1/3}(e^{-i\pi}\chi) +\mathrm{J}_{-1/3}(e^{i\pi}\chi)\,\mathrm{J}_{-1/3}(e^{-i\pi}\chi)
\nonumber\\
&\;
+\mathrm{J}_{1/3}(e^{i\pi}\chi)\,\mathrm{J}_{-1/3}(e^{-i\pi}\chi) +\mathrm{J}_{1/3}(e^{-i\pi}\chi)\,\mathrm{J}_{-1/3}(e^{i\pi}\chi)
\nonumber\\[5pt]
&=\big[\mathrm{J}_{1/3}(e^{i\pi}\chi)
+\mathrm{J}_{-1/3}(e^{i\pi}\chi)
\big]
\nonumber
\\
&\qquad\quad
\times\big[
\mathrm{J}_{1/3}(e^{-i\pi}\chi)
+\mathrm{J}_{-1/3}(e^{-i\pi}\chi)
\big]
\nonumber
\end{align}
and 
rewrite Eq.\,(\ref{eq204}) in a multiplicative form:
\begin{align}
&\textstyle
\mathcal{R}(\mathcal{A})
\nonumber\\
&\textstyle
=\frac{
\frac{9}{4\pi^2\mathcal{A}}}{\left[\mathrm{J}_{1/3}(e^{i\pi}\chi)
+\mathrm{J}_{-1/3}(e^{i\pi}\chi)
\right]\left[
\mathrm{J}_{1/3}(e^{-i\pi}\chi)
+\mathrm{J}_{-1/3}(e^{-i\pi}\chi)
\right]}.
\label{eq205}
\end{align}

For the analytic calculation of the population-mean potential $v$ in Sec.\,\ref{ssec33:MV}, we will need the quadrature form of solution for $r$~\cite{Hirsch-Huberman-Scalapino-1982,Lindner-Longtin-Bulsara-2003} derived directly from the time-independent solution of the Fokker--Planck equation (\ref{eq103}).
For a time-independent probability density $P(V)$, Eq.\,(\ref{eq103}) can be integrated once over $V$. This yields the probability flux, which equals the firing rate $r$~\cite{Hirsch-Huberman-Scalapino-1982,Lindner-Longtin-Bulsara-2003}:
\begin{equation}
r=(A+V^2)P-D\frac{\partial P}{\partial V}\,,
\label{eq201}
\end{equation}
this equation can be solved with
\begin{equation}
P(V)=\frac{r}{D}\int\nolimits_V^{+\infty}\mathrm{d}U\,e^{-\frac{A}{D}(U-V)-\frac{U^3-V^3}{3D}}.
\label{eq202}
\end{equation}
The normalization condition $\int_{-\infty}^{+\infty}P(V)\,\mathrm{d}V=1$ determines $r$. With the substitution $U\to V+y$, we find
\begin{align}
1 =\int\limits_{-\infty}^{+\infty}\mathrm{d}V\,P(V) &=\frac{r}{D}\int\limits_{-\infty}^{+\infty}\mathrm{d}V \int\limits_{0}^{+\infty}\mathrm{d}y\,e^{\frac{-(A+V^2)y-Vy^2-\frac{y^3}{3}}{D}}
\nonumber\\
&
=\frac{r}{D}\int\limits_{0}^{+\infty}\mathrm{d}y\,e^{\frac{-Ay-\frac{y^3}{12}}{D}}  \int\limits_{-\infty}^{+\infty}\mathrm{d}V\,e^{\frac{-y(V+\frac{y}{2})^2}{D}}
\nonumber\\
&
=\frac{r}{D^{1/3}}\int\limits_{0}^{+\infty}\frac{\sqrt{\pi}\mathrm{d}Y}{\sqrt{Y}}  e^{-\mathcal{A}Y-\frac{Y^3}{12}},
\label{eq203}
\end{align}
where the integral over $V$ is evaluated as an integral of a Gaussian distribution, and
we have introduced notation
$$Y=y/D^{1/3}.$$

\subsection{Population-mean membrane potential for\\ a time-independent macroscopic state}
\label{ssec33:MV}
For the distribution (\ref{eq202}), we can use the definition of $r$ and $\mathcal{R}(\mathcal{A})$ via the integral~(\ref{eq203}) to calculate the population-mean membrane potential:
\begin{align}
v&=\mathrm{P.V.}\int\limits_{-\infty}^{+\infty}\mathrm{d}V\,VP(V)
\nonumber\\
&=\frac{r}{D}\,\mathrm{P.V.}\int\limits_{-\infty}^{+\infty}\mathrm{d}V
\int\limits_{0}^{+\infty}\mathrm{d}y\,V e^{\frac{-(A+V^2)y-Vy^2-\frac{y^3}{3}}{D}}
\nonumber\\
&
=\frac{r}{D}\int\limits_{0}^{+\infty}\mathrm{d}y\,e^{\frac{-Ay-\frac{y^3}{12}}{D}}  \int\limits_{-\infty}^{+\infty}\mathrm{d}V\,Ve^{\frac{-y(V+\frac{y}{2})^2}{D}}
\nonumber\\
&
=-\frac{r}{2}\int\limits_{0}^{+\infty}\mathrm{d}Y\sqrt{\pi Y}  e^{-\mathcal{A}Y-\frac{Y^3}{12}}
\nonumber
\\
&
=\frac{r}{2}\frac{\mathrm{d}}{\mathrm{d}\mathcal{A}}
\int\limits_{0}^{+\infty}\frac{\sqrt{\pi}\mathrm{d}Y}{\sqrt{Y}} e^{-\mathcal{A}Y-\frac{Y^3}{12}}
=\frac{D^{1/3}\mathcal{R}(\mathcal{A})}{2} \frac{\mathrm{d}}{\mathrm{d}\mathcal{A}}\frac{1}{\mathcal{R}(\mathcal{A})}
\nonumber
\\
&
=\frac{D^{1/3}}{2}\frac{\mathrm{d}}{\mathrm{d}\mathcal{A}}\ln\frac{1}{\mathcal{R}(\mathcal{A})}\,,
\label{eq206}
\end{align}
where ``P.V.'' indicates a principal value integral, and
the integral over $V$ is evaluated as the average of $V$ for a shifted Gaussian distribution.
Further, we introduce notation
\[
v=D^{1/3}\mathcal{V}(\mathcal{A})
\]
and calculate function $\mathcal{V}(\mathcal{A})$ from (\ref{eq206}) and (\ref{eq205}). Using the relations $\mathrm{J}_\alpha^\prime(\chi)=\pm\mathrm{J}_{\alpha\mp1}(\chi)\mp\alpha\mathrm{J}_\alpha(\chi)/\chi$ and picking the $(\alpha\pm1)$-case for $\alpha=\mp1/3$, we can calculate
\begin{align}
\mathcal{V}(\mathcal{A}) &
=\frac{1}{2}\frac{\mathrm{d}}{\mathrm{d}\mathcal{A}}\ln\frac{1}{\mathcal{R}(\mathcal{A})}
\nonumber\\[5pt]
&
=\frac{1}{2\mathcal{A}}+\frac{\sqrt{\mathcal{A}}}{2}\left[
e^{i\pi}\frac{\mathrm{J}_{1/3}^\prime(e^{i\pi}\chi) +\mathrm{J}_{-1/3}^\prime(e^{i\pi}\chi)}
{\mathrm{J}_{1/3}(e^{i\pi}\chi)+\mathrm{J}_{-1/3}(e^{i\pi}\chi)}
\right.
\nonumber\\
&\qquad\quad\left.{}
+e^{-i\pi}\frac{\mathrm{J}_{1/3}^\prime(e^{-i\pi}\chi) +\mathrm{J}_{-1/3}^\prime(e^{-i\pi}\chi)}
{\mathrm{J}_{1/3}(e^{-i\pi}\chi)+\mathrm{J}_{-1/3}(e^{-i\pi}\chi)}\right]
\label{eq207}\\[5pt]
&
=\frac{\sqrt{\mathcal{A}}}{2}\left[
\frac{\mathrm{J}_{2/3}(e^{i\pi}\chi)-\mathrm{J}_{-2/3}(e^{i\pi}\chi)}
{\mathrm{J}_{1/3}(e^{i\pi}\chi)+\mathrm{J}_{-1/3}(e^{i\pi}\chi)}
\right.
\nonumber\\
&\qquad\qquad\;\;\left.{}
 +\frac{\mathrm{J}_{2/3}(e^{-i\pi}\chi) -\mathrm{J}_{-2/3}(e^{-i\pi}\chi)}
{\mathrm{J}_{1/3}(e^{-i\pi}\chi)+\mathrm{J}_{-1/3}(e^{-i\pi}\chi)}\right].
\label{eq208}
\end{align}

For real $\mathcal{A}>0$, we find $\chi>0$ and the second term in the brackets of (\ref{eq208}) is obviously the complex conjugate of the first terms. For $\mathcal{A}<0$, we find $\chi=e^{i3\pi/2}|\chi|$ and Eq.\,(\ref{eq208}) can be rewrite as
\begin{align}
&\mathcal{V}(\mathcal{A})
=\frac{e^{i\pi/2}\sqrt{|\mathcal{A}|}}{2}\left[
\frac{\mathrm{J}_{2/3}(e^{i5\pi/2}|\chi|)-\mathrm{J}_{-2/3}(e^{i5\pi/2}|\chi|)}
{\mathrm{J}_{1/3}(e^{i5\pi/2}|\chi|)+\mathrm{J}_{-1/3}(e^{i5\pi/2}|\chi|)}\right.
\nonumber\\
&\qquad\qquad\qquad\left.{}+
\frac{\mathrm{J}_{2/3}(e^{i\pi/2}|\chi|) -\mathrm{J}_{-2/3}(e^{i\pi/2}|\chi|)}
{\mathrm{J}_{1/3}(e^{i\pi/2}|\chi|)+\mathrm{J}_{-1/3}(e^{i\pi/2}|\chi|)}\right]
\nonumber\\[5pt]
&\quad
=\frac{\sqrt{|\mathcal{A}|}}{2}\left[
e^{-i\pi/2}\frac{\mathrm{J}_{2/3}(e^{-i\pi/2}|\chi|) -\mathrm{J}_{-2/3}(e^{-i\pi/2}|\chi|)}
{\mathrm{J}_{1/3}(e^{-i\pi/2}|\chi|) +\mathrm{J}_{-1/3}(e^{-i\pi/2}|\chi|)}\right.
\nonumber\\
&
\qquad\quad\left.{}+
e^{i\pi/2}\frac{\mathrm{J}_{2/3}(e^{i\pi/2}|\chi|) -\mathrm{J}_{-2/3}(e^{i\pi/2}|\chi|)}
{\mathrm{J}_{1/3}(e^{i\pi/2}|\chi|)+\mathrm{J}_{-1/3}(e^{i\pi/2}|\chi|)}\right],
\label{eq208neg}
\end{align}
where the second term is again the complex conjugate of the first term. Hence, for real $\mathcal{A}$, for which the physical problem is solved, we can express Eqs.\,(\ref{eq208}) and (\ref{eq208neg}) in a compact form:
\begin{align}
\mathcal{V}(\mathcal{A})&=-\mathrm{Im}\mathcal{W}(\mathcal{A})\,,
\label{eq209}
\\[5pt]
\mathcal{W}(\mathcal{A})&=\frac{\sqrt{\mathcal{A}}}{i}
\frac{\mathrm{J}_{2/3}(e^{i\pi}\chi)-\mathrm{J}_{-2/3}(e^{i\pi}\chi)}
{\mathrm{J}_{1/3}(e^{i\pi}\chi)+\mathrm{J}_{-1/3}(e^{i\pi}\chi)}\,.
\label{eq210}
\end{align}

\subsection{Montbri\'o--Paz\'o--Roxin order parameter\\ (the first pseudocumulant)}
\label{ssec34:MPR}
Let us now find the intimate connection between the expressions for $\mathcal{R}(\mathcal{A})$ (\ref{eq205}) and $\mathcal{V}(\mathcal{A})$ (\ref{eq208}) [or (\ref{eq209})].
We consider $\mathrm{Re}\mathcal{W}$ for positive real $\mathcal{A}>0$ (real $\chi>0$), transform it back to a shape similar to (\ref{eq207}), converting
from $\mathrm{J}_{\pm2/3}$ back to $\mathrm{J}_{\pm1/3}$ and their derivatives,
and use the property $\mathrm{J}_\alpha^\prime(e^{in\pi}Z)=e^{i(\alpha-1)n\pi}\mathrm{J}_\alpha^\prime(Z)$\,:
\begin{widetext}
\begin{align}
\mathrm{Re}\,\mathcal{W}(\mathcal{A})&=
\frac{\sqrt{\mathcal{A}}}{2i}\left[
\frac{\mathrm{J}_{2/3}(e^{i\pi}\chi)-\mathrm{J}_{-2/3}(e^{i\pi}\chi)}
{\mathrm{J}_{1/3}(e^{i\pi}\chi)+\mathrm{J}_{-1/3}(e^{i\pi}\chi)}
-\frac{\mathrm{J}_{2/3}(e^{-i\pi}\chi) -\mathrm{J}_{-2/3}(e^{-i\pi}\chi)}
{\mathrm{J}_{1/3}(e^{-i\pi}\chi)+\mathrm{J}_{-1/3}(e^{-i\pi}\chi)}\right]
\nonumber
\\
&=\frac{\sqrt{\mathcal{A}}}{2}\left[
-\frac{\mathrm{J}_{1/3}^\prime(e^{i\pi}\chi) +\mathrm{J}_{-1/3}^\prime(e^{i\pi}\chi)}
{\mathrm{J}_{1/3}(e^{i\pi}\chi)+\mathrm{J}_{-1/3}(e^{i\pi}\chi)}
+\frac{\mathrm{J}_{1/3}^\prime(e^{-i\pi}\chi) +\mathrm{J}_{-1/3}^\prime(e^{-i\pi}\chi)}
{\mathrm{J}_{1/3}(e^{-i\pi}\chi)+\mathrm{J}_{-1/3}(e^{-i\pi}\chi)}\right]
\nonumber
\\
&
=
\frac{-\frac{\sqrt{3}}{2}\sqrt{\mathcal{A}}
\left[\mathrm{J}_{1/3}(\chi)\,\mathrm{J}_{-1/3}^\prime(\chi) -\mathrm{J}_{-1/3}(\chi)\,\mathrm{J}_{1/3}^\prime(\chi)\right]
}{\left[\mathrm{J}_{1/3}(e^{i\pi}\chi)
+\mathrm{J}_{-1/3}(e^{i\pi}\chi)
\right]\left[
\mathrm{J}_{1/3}(e^{-i\pi}\chi)
+\mathrm{J}_{-1/3}(e^{-i\pi}\chi)
\right]}.
\label{eq211}
\end{align}
\end{widetext}
Using Abel's identity in the numerator,
$$
\mathrm{J}_{\alpha}(\chi)\mathrm{J}_{-\alpha}^\prime(\chi) -\mathrm{J}_{-\alpha}(\chi)\mathrm{J}_{\alpha}^\prime(\chi)
=-\frac{2\sin{\alpha\pi}}{\pi\chi}
$$
with $\alpha=1/3$, we can rewrite Eq.\,(\ref{eq211}) as $\pi\mathcal{R}(\mathcal{A})$, where $\mathcal{R}(\mathcal{A})$ is given by Eq.\,(\ref{eq205}).
Similarly to the case of $\mathcal{V}(\mathcal{A})$, for real $\mathcal{A}<0$ and $\chi=e^{i3\pi/2}|\chi|$, we can again find that $\mathrm{Re}\mathcal{W}(\mathcal{A})=\pi\mathcal{R}(\mathcal{A})$.

\begin{figure}[!b]
\center{
\includegraphics[width=0.99\columnwidth]%
 {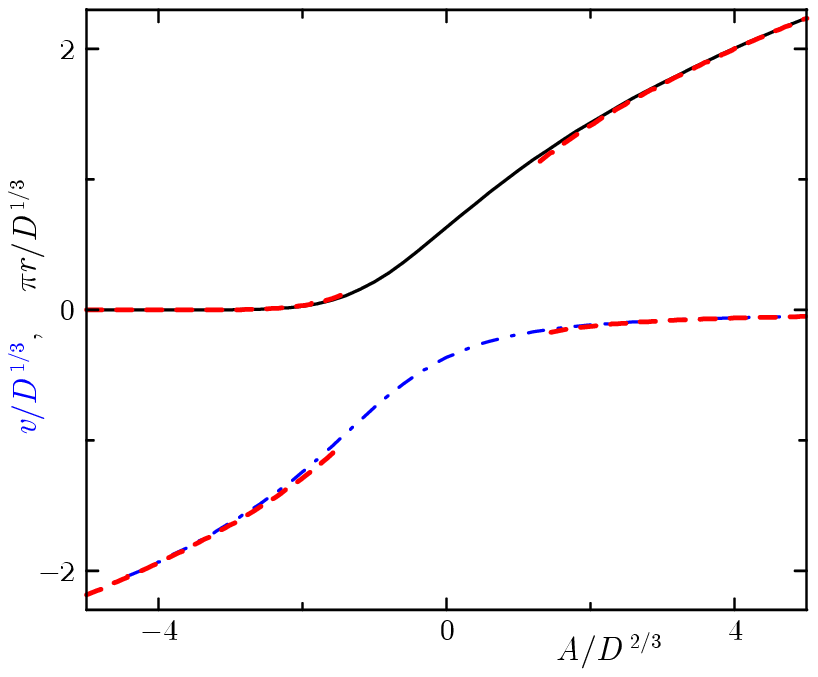}
}
\caption{The firing rate $r$ of the quadratic integrate-and-fire neuron~(\ref{eq101}) (black solid line) and the population-mean potential $v$ (blue dash-dotted line) are plotted for the analytical solution (\ref{eq212}) and (\ref{eq210}) [or (\ref{eq204}) and (\ref{eq208})]. Red dashed lines present the asymptotic laws~(\ref{eq21r}) and (\ref{eq21v}).
}
  \label{fig1}
\end{figure}

Summarizing, for real $\mathcal{A}$, we can analytically calculate the MPR order parameter $W=\pi r-iv$ \cite{Montbrio-Pazo-Roxin-2015}, which is also the first pseudocumulant~\cite{Goldobin-Volo-Torcini-2021}:
\begin{align}
&W=\pi r-iv=D^{1/3}\mathcal{W}(\mathcal{A})\,,
\label{eq212}\\
&\mathcal{W}(\mathcal{A})=\pi\mathcal{R}(\mathcal{A})-i\mathcal{V}(\mathcal{A})\,,
\label{eq213}
\end{align}
where $\mathcal{W}(\mathcal{A})$ is given by Eq.\,(\ref{eq210}), see Fig.\,\ref{fig1}.

The asymptotic laws for (\ref{eq204}) and (\ref{eq208}) are practically useful~\cite{Lindner-Longtin-Bulsara-2003,Volo-etal-2022};
\begin{align}
r&\approx
\left\{
\begin{array}{cc}
\displaystyle
\frac{\sqrt{|A|}}{\pi}\exp\left(-\frac{4|A|^{3/2}}{3D}\right) &
\displaystyle
\mbox{for }
\frac{A}{D^{2/3}}\to -\infty , \\[7pt]
\displaystyle
\frac{\sqrt{A}}{\pi}\,&
\displaystyle
\mbox{for }
\frac{A}{D^{2/3}}\to+\infty ;
\end{array}
\right.
\label{eq21r}
\\
v&\approx
\left\{
\begin{array}{cc}
\displaystyle
-\sqrt{|A|}+\frac{D}{4|A|} &
\displaystyle
\mbox{ for }
\frac{A}{D^{2/3}}\to -\infty , \\[7pt]
\displaystyle
-\frac{D}{4A}\,&
\displaystyle
\mbox{ for }
\frac{A}{D^{2/3}}\to+\infty .
\end{array}
\right.
\label{eq21v}
\end{align}
The accuracy of these laws can be assessed from Fig.\,\ref{fig1}.

\begin{figure}[!t]
\centerline{
\includegraphics[width=0.47\textwidth]%
 {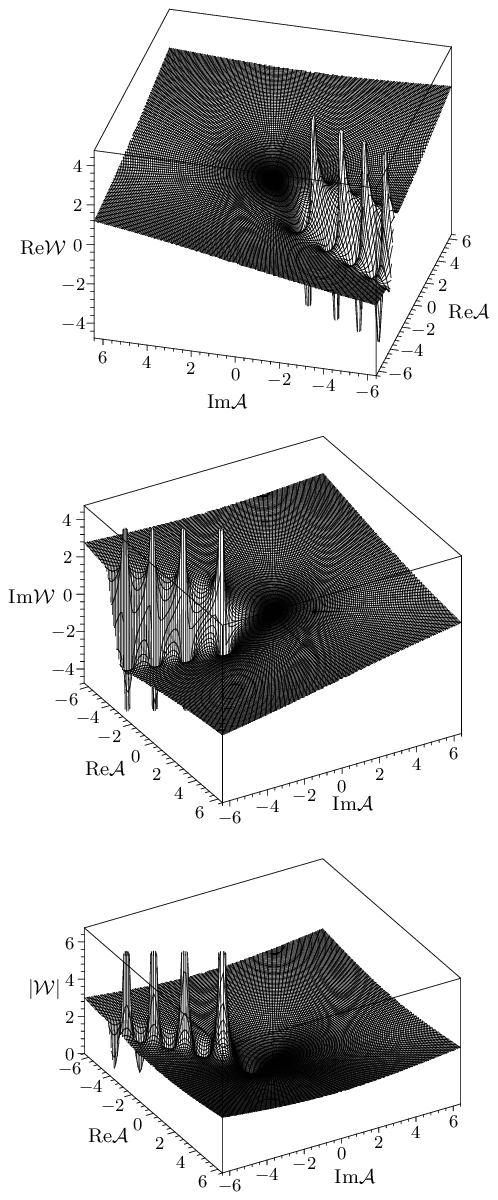}
}
\caption{Analytic continuation of $\mathcal{W}(\mathcal{A})$ into complex plane.
}
  \label{fig2}
\end{figure}

\begin{figure}[!t]
\centerline{
\includegraphics[width=0.99\columnwidth]%
 {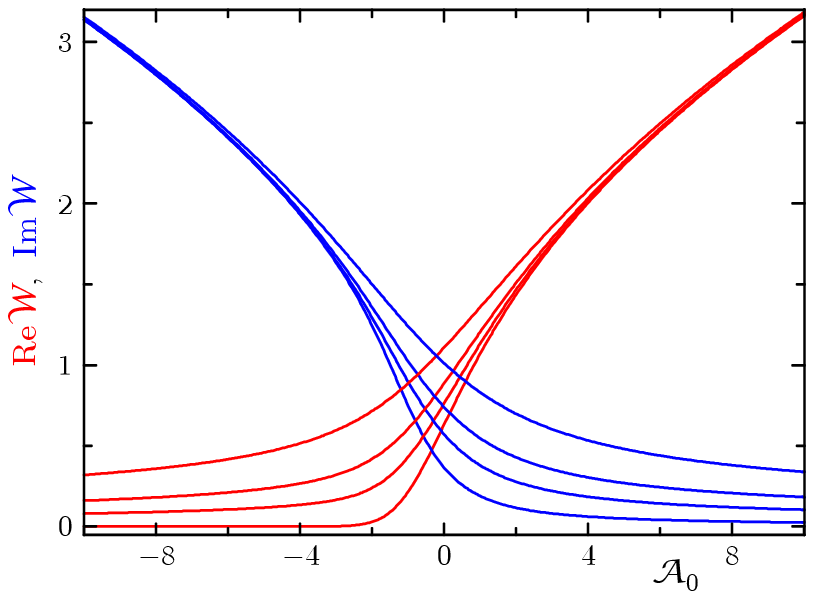}
}
\caption{The firing rate $\langle{r}\rangle$ of the quadratic integrate-and-fire neuron~(\ref{eq101}) (red lines) and the population-mean potential $\langle{v}\rangle$ (blue lines) are given by the MPR order parameter $\pi\langle{r}\rangle-i\langle{v}\rangle=\langle{W}\rangle =D^{1/3}\mathcal{W}(\mathcal{A}_0+\Delta_\mathcal{A})$ for the time-independent macroscopic state of a homogeneous population and heterogeneous ones with a Lorentzian distribution of $\mathcal{A}$ centered at $\mathcal{A}_0$; half width at half maximum of the distribution $\Delta_\mathcal{A}=0$, $0.5$, $1$, $2$ (from bottom to top).
}
  \label{fig3}
\end{figure}

\section{Heterogeneous populations}
\label{sec4:Heter}
For a heterogeneous population of QIFs with a Lorentzian distribution of $A$ centered at $A_0$ with a half width at half maximum $\Delta_A$, $L_A(A)=\Delta_A/\{\pi[\Delta_A^2+(A-A_0)^2]\}$, the population mean values of the firing rate and membrane potential are determined by the weighted integral over real $A$:
\begin{align}
\pi\langle r\rangle-i\langle v\rangle
&=\int\limits_{-\infty}^{+\infty} D^{1/3}\mathcal{W}(\mathcal{A})L_\mathcal{A}(\mathcal{A})\,\mathrm{d}\mathcal{A}\,,
\label{eq301}
\end{align}
where
\begin{align}
&L_\mathcal{A}(\mathcal{A})
=\frac{\pi^{-1}\Delta_\mathcal{A}}{\Delta_\mathcal{A}^2+(\mathcal{A}-\mathcal{A}_0)^2}
\nonumber\\
&
=\frac{1}{2\pi i}\left[\frac{1}{\mathcal{A}-(\mathcal{A}_0+i\Delta_\mathcal{A})}
 -\frac{1}{\mathcal{A}-(\mathcal{A}_0-i\Delta_\mathcal{A})}\right],
\label{eq302}
\end{align}
$\Delta_\mathcal{A}=\Delta_A/D^{2/3}$, $\mathcal{A}_0=A_0/D^{2/3}$.

We need to evaluate
$
\left\langle\mathcal{W}\right\rangle=\int\nolimits_{-\infty}^{+\infty} \mathcal{W}(\mathcal{A})L_\mathcal{A}(\mathcal{A})\,\mathrm{d}\mathcal{A}.
$
Considering the analytic continuation of $\mathcal{W}(\mathcal{A})$ (\ref{eq210}) into the complex plane of $\mathcal{A}$, we note that although some functions constituting $\mathcal{W}(\mathcal{A})$ are multi-valued, their combination forms a single-valued (single-sheet) analytic function (see Fig.\,\ref{fig2}). Indeed, $\mathcal{W}(e^{i2\pi}\mathcal{A})=\mathcal{W}(\mathcal{A})$, while $\sqrt{e^{i2\pi}\mathcal{A}}=e^{i\pi}\sqrt{\mathcal{A}}$ and $\mathrm{J}_{\pm1/3}[e^{i\pi}(2/3)(e^{i2\pi}\mathcal{A})^{3/2}]=e^{\pm i\pi}\mathrm{J}_{\pm1/3}[e^{i\pi}(2/3)\mathcal{A}^{3/2}]$. In Fig.\,\ref{fig2} and Eq.\,(\ref{eq210}), one can see that this function is smooth at $\mathcal{A}\to0$ and possesses an infinite chain of poles at $\mathcal{A}=e^{i4\pi/3}(3\gamma_n/2)^{2/3}$, where $\gamma_n$ with $n=1,2,3,...$ are the roots of
$$
\mathrm{J}_{1/3}(\gamma)+\mathrm{J}_{-1/3}(\gamma)=0.
$$
This sum of Bessel functions can be zero only for positive real $\gamma$. Specifically, the first six $\gamma_1=2.3834466...$, $\gamma_2=5.5101956...$, $\gamma_3=8.6473577...$, $\gamma_4=11.786843...$, $\gamma_5=14.927207...$, $\gamma_6=18.067995...$; for large $n$, $\gamma_{n\gg1}\approx\pi(n-1/4)$\,. Further, for large $|\mathcal{A}|$ and away from the chain of poles (see the bottom panel of Fig.\,\ref{fig2}), we find $|\mathcal{W}(\mathcal{A})|\propto\sqrt{|\mathcal{A}|}$.
Therefore, we can close the contour over the upper arc at infinity; indeed, for this contour,
\begin{align}
\oint\mathcal{W}(\mathcal{A})L_\mathcal{A}(\mathcal{A})\,\mathrm{d}\mathcal{A}=
\int\limits_{-\infty}^{+\infty}\! \mathcal{W}(\mathcal{A})L_\mathcal{A}(\mathcal{A})\,\mathrm{d}\mathcal{A}
\quad
\nonumber\\
{}+\lim_{B\to+\infty}\int\limits_0^\pi\! \mathcal{W}(Be^{i\phi})L_\mathcal{A}(Be^{i\phi})\,Be^{i\phi}i\mathrm{d}\phi\,,
\nonumber
\end{align}
where the latter integral tends to zero:
\begin{align}
&\left|\lim_{B\to+\infty}\int\limits_0^\pi\!\mathcal{W}(Be^{i\phi})L_\mathcal{A}(Be^{i\phi})\,Be^{i\phi}i\mathrm{d}\phi\right| \leq
\nonumber\\
& \lim_{B\to+\infty}B\int\limits_0^\pi\!\!|\mathcal{W}(Be^{i\phi})||L_\mathcal{A}(Be^{i\phi})|\mathrm{d}\phi
=\lim_{B\to+\infty}\frac{\Delta_\mathcal{A}}{\sqrt{B}}=0.
\qquad\qquad
\nonumber
\end{align}
By virtue of the Residue theorem, the integral
of an analytic function over a closed path equals to the sum of the residues at the poles inside it; there is only one such pole in the upper half-plane and it is located at $\mathcal{A}_0+i\Delta_\mathcal{A}$ [see Eq.\,(\ref{eq302})]. Finally,
\begin{align}
\left\langle\mathcal{W}\right\rangle
&
=\oint\frac{\mathcal{W}(\mathcal{A})\,\mathrm{d}\mathcal{A}}
{2\pi i [\mathcal{A}-(\mathcal{A}_0+i\Delta_\mathcal{A})]}
=\mathcal{W}(\mathcal{A}_0+i\Delta_\mathcal{A})\,.
\label{eq303}
\end{align}
In Fig.\,\ref{fig3} one can see the dependencies of $\langle{r}\rangle$ and $\langle{v}\rangle$ on $\mathcal{A}_0$ for heterogeneous populations.

Summarizing, the analytic continuation of the solution enables us to derive an algebraic self-consistency problem for a population of elements with a Lorentzian distribution of parameters~\cite{Yakubovich-1969,Rabinovich-Trubetskov-1989,Crawford-1994,Ott-Antonsen-2008}.
Using result~(\ref{eq303}), we can extend these findings to other parameter distributions~\cite{Klinshov-Kirillov-Nekorkin-2021}.
We emphasize that the Residue theorem can be used for $\mathcal{W}$ (\ref{eq210}) but was not for $\mathcal{R}(\mathcal{A})$ (\ref{eq204}), as the latter has two chains of poles, one in each half-plane.

An important technical note: computer calculations of Eq.\,(\ref{eq204}) with negative $\mathcal{A}$ for the $\chi$-branch and with positive $\mathcal{A}$ for the $\chi_-$-branch ({\it i.e.}, with an interchanged choice of branches), as well as calculations of Eq.\,(\ref{eq210}) for purely real $\mathcal{A}$, require careful monitoring to avoid sporadic inconsistencies with the choice of sheets of multivalued complex functions.
This is why we provide explicit expressions (\ref{eq204}) for positive and negative $\mathcal{A}$, which involve purely real calculations. Meanwhile, this issue does not arise for nonzero imaginary part of $\mathcal{A}$ in Eq.\,(\ref{eq210}); calculations remain reliable even for as small $\mathrm{Im}\mathcal{A}$ as $10^{-15}$.

\section{Mean-field models}
\label{sec5:mean-field}
\subsection{Networks with electrical and chemical synapses}
\label{ssec51:MFM1}
The first example we consider concerns the interplay between electrical and chemical synapses and the effect of Gaussian fluctuations on it.
The nontrivial role of electrical synapses was studied within the framework of the exact mean-field model, given by an amended MPR theory for QIFs with asymmetric spiking~\cite{Montbrio-Pazo-2020}.
For this theory and symmetric spiking, as we will see in Sec.\,\ref{ssec51a:MFM1a}, the mean-field impact of electrical synapses is equivalent to a mere decrease of the chemical synapse heterogeneity. Using analytical results (\ref{eq210}) and (\ref{eq303}), we can study electrical and chemical synapses in networks with fluctuations, beyond the MPR theory~\cite{Montbrio-Pazo-Roxin-2015}, and observe that the impact of electrical synapses is generally more sophisticated than a mere recalibration of heterogeneity parameters.
Without loss of generality, we will consider only symmetric spiking here.
Since the macroscopic dynamics of networks with partial connectivity can often be reproduced in networks with global coupling~\cite{Clusella-Montbrio-2024}, we consider a network of $N$ QIFs (\ref{eq101})--(\ref{eq102}) with
\begin{align}
A_j=\eta_j+J_j  S(t)+a\left[v(t)-V_j\right]  \,,
\label{eq:MFAj}
\\
S(t)=\frac{1}{N}\sum_{j=1}^N\sum_n\delta(t-t_j^{(n)})\,,\quad
\end{align}
where $\eta_j$ and $J_j$ represent external currents and chemical synaptic couplings with Lorentzian distributions centered at $\eta_0$ and $J_0$, with half-width $\Delta_\eta$ and $\Delta_J$, respectively;
$S(t)$ is the net synaptic activity, $t_j^{(n)}$ is the time of the $n$-th firing event of the $j$-th neuron; the population-mean membrane potential is $v(t)=N^{-1}\sum_{j=1}^NV_j(t)$; and
$a$ is the strength of electrical coupling~\cite{Montbrio-Pazo-2020}.
The chemical synaptic coupling is excitatory for $J_j>0$ and inhibitory for $J_j<0$.

\subsubsection{No Gaussian noise case ($D=0$)}\label{ssec51a:MFM1a}
For the case of $D=0$ and symmetric spiking, one can derive an exact dynamic equation for $W=\pi r(t)-iv(t)$~\cite{Montbrio-Pazo-2020} and use it for comparison with the case $D\ne0$ to better understand the influence of noise.
With the transformation $U_j=V_j-a/2$ and $u=v-a/2$, one can eliminate the linear-in-$V_j$ term from (\ref{eq101}) with $A_j$ given by (\ref{eq:MFAj}).
For the thermodynamic limit $N\gg1$, where the firing rate $r=\langle S(t)\rangle$, the MPR theory~\cite{Montbrio-Pazo-Roxin-2015} {\it or} the pseudocumulant expansion~\cite{Goldobin-Volo-Torcini-2021} yields the complex equation of dynamics:
\begin{equation}
\dot{W}_u=iW_u^2-iA_u(t)+\Delta_{A}(t)\,,
\label{eqSM3A01}
\end{equation}
where
\begin{align}
W_u&=\pi r(t)-i u(t) \, ,
\nonumber\\
A_u(t)&=\eta_0+\frac{a^2}{4}+au(t)+J_0r(t) \, ,
\nonumber\\
\Delta_{A}(t)&=\Delta_\eta+\Delta_J r(t) \, .
\nonumber
\end{align}
Considering the real and imaginary parts of Eq.\,(\ref{eqSM3A01}) separately, we obtain
\begin{align}
\dot{r}&=\frac{\Delta_\eta}{\pi} +\left(2v +\frac{\Delta_J}{\pi}-a\right)r\,,
\label{eqSM3A02}\\
\dot{v}&=v^2+\eta_0-\pi^2r^2+J_0r\,.
\label{eqSM3A03}
\end{align}
Here, the electrical coupling strength $a$ and the chemical synapse heterogeneity $\Delta_J$ appear only in the form of the combination
\begin{equation}
a_J=a-\frac{1}{\pi}\Delta_J\,,
\label{eq:MFaJ}
\end{equation}
meaning that, on the mean-field level, the chemical synapse heterogeneity effectively acts as a negative electrical coupling. Notice, physically, the electrical coupling $a$ must be positive since it corresponds to the Ohmic leakage current between adjacent membranes, with a nonnegative resistance (conductance) coefficient. In turn, the heterogeneity $\Delta_J$ is also strictly nonnegative.

For time-independent states, Eqs.\,(\ref{eqSM3A02}) and (\ref{eqSM3A03}) yield quartic equations for $r$ or $v$, but the solutions can be expressed in parametric form as follows. For the dependencies of $v$ and on $\eta_0$ and $r$ on $\eta_0$:
\begin{align}
\eta_0&=\frac{\Delta_\eta^2}{(2v-a_J)^2}-v^2+\frac{\Delta_\eta J_0/\pi}{2v-a_J}\,,
\label{eqSM3A04}\\
\eta_0&=\pi^2r^2-J_0r-\frac{a_J^2}{4}+\frac{\Delta_\eta a_J}{2\pi r} -\frac{\Delta_\eta^2}{4\pi^2r^2}\,,
\label{eqSM3A05}
\end{align}
parameterized by $v\in(-\infty,a_J/2)$ and $r\in(0,+\infty)$, respectively.

For time-independent states, we can also determine the bistability domain bounded by tangential bifurcations. By setting $\dot{r}=\dot{v}=0$ in Eqs.\,(\ref{eqSM3A02}) and (\ref{eqSM3A03}), and taking differentials $\mathrm{d}r\ne0$ and $\mathrm{d}v\ne0$ under the condition $\mathrm{d}\eta_0=\mathrm{d}J_0=0$, we obtain
\begin{align}
J_{0\ast}&=2\pi^2r_\ast
 +\frac{\Delta_\eta}{\pi r_\ast^2}\left(\frac{\Delta_\eta}{2\pi r_\ast}-\frac{a_J}{2}\right),
\label{eqSM3A06}\\
\eta_{0\ast}&=
 \pi^2r_\ast^2-J_{0\ast}r_\ast
 -\left(\frac{\Delta_\eta}{2\pi r_\ast}-\frac{a_J}{2}\right)^2,
\label{eqSM3A07}
\end{align}
which is the bistability domain boundary $(\eta_{0\ast},J_{0\ast})$ parameterized by $r_\ast\in(0,+\infty)$.

\begin{figure}[!t]
\centerline{
\includegraphics[width=0.99\columnwidth]%
 {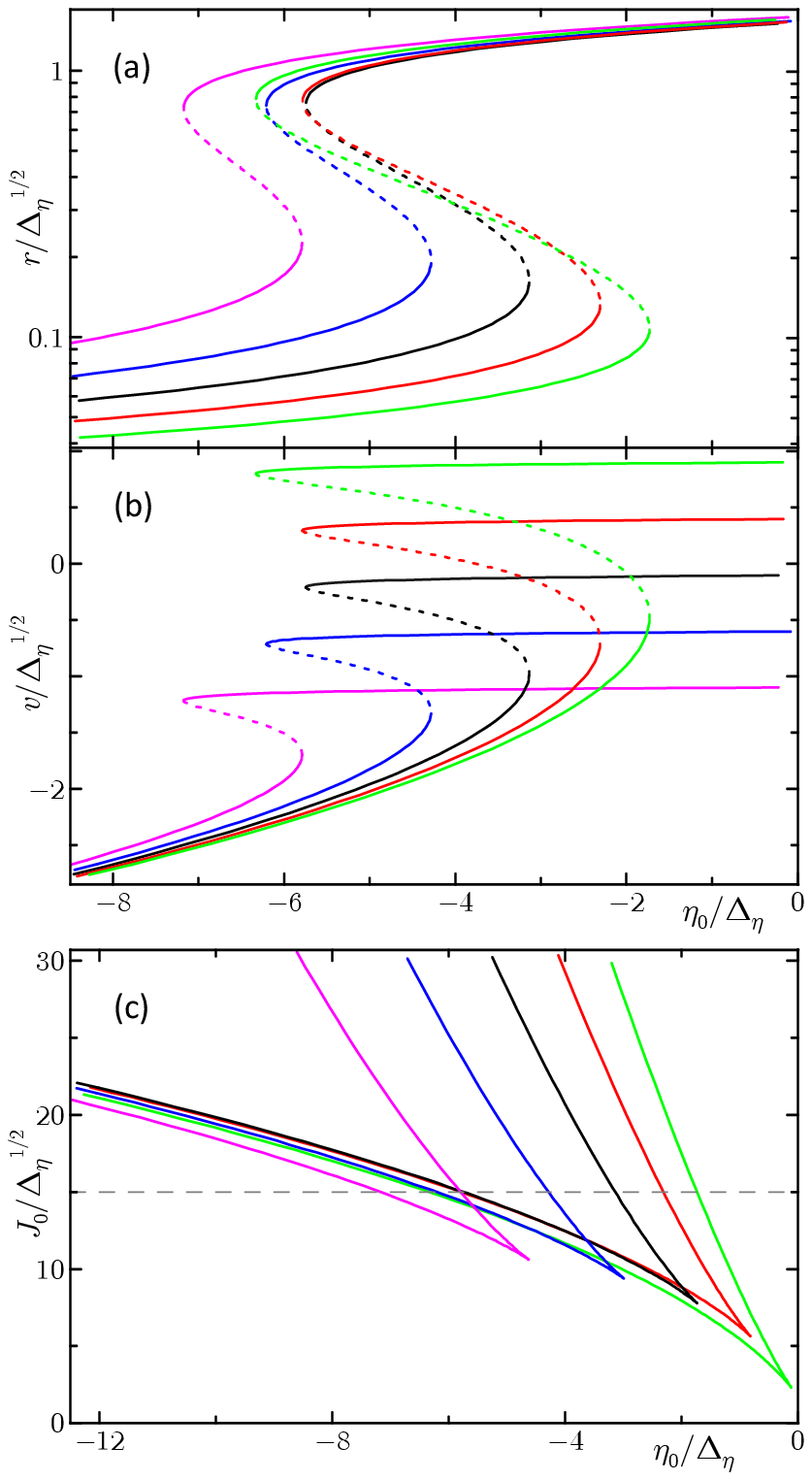}
}
\caption{Population firing rate $r$ (a) and mean potential $v$ (b) are plotted {\it versus} $\eta_0$ for a network with electrical and chemical synapses at $D=0$ [Eqs.\,(\ref{eqSM3A04}) and (\ref{eqSM3A05})], $J_0/\Delta_\eta^{1/2}=15$ [grey dashed line in panel~(c)], $a_J/\Delta_\eta^{1/2}=-2$ (magenta), $-1$ (blue), $0$ (black), $1$ (red), $2$ (green); dotted lines: the unstable branch of solution.
The hysteretic transition domain is within the wedge in panel (c) given by Eqs.\,(\ref{eqSM3A06}) and (\ref{eqSM3A07}) for $a_J/\Delta_\eta^{1/2}=-2$, $-1$, $0$, $1$, $2$ with the same color coding as in panels (a,b).
}
  \label{fig4}
\end{figure}

The dependencies of the population firing rate and mean potential {\it versus} $\eta_0$ are presented in Fig.\,\ref{fig4}(a,b) for the domain of bistability between the low- and high-firing-rate regimes. These regimes are characterized by negative and positive $A_0$, respectively, which correspond to the predominance of excitable states and states of periodic spiking for individual QIFs. In panel~(c) of Fig.\,\ref{fig4}, the boundaries of the bistability domain are plotted for different $a_J$.
Negative values of $a_J$ (\ref{eq:MFaJ}) correspond to a stronger synaptic heterogeneity $\Delta_J$ prevailing over the electrical coupling $a$, while positive values correspond to a stronger coupling prevailing over the heterogeneity impact.

\subsubsection{Effect of Gaussian noise ($D\ne0$)}
\begin{figure}[!t]
\centerline{
\includegraphics[width=0.99\columnwidth]%
 {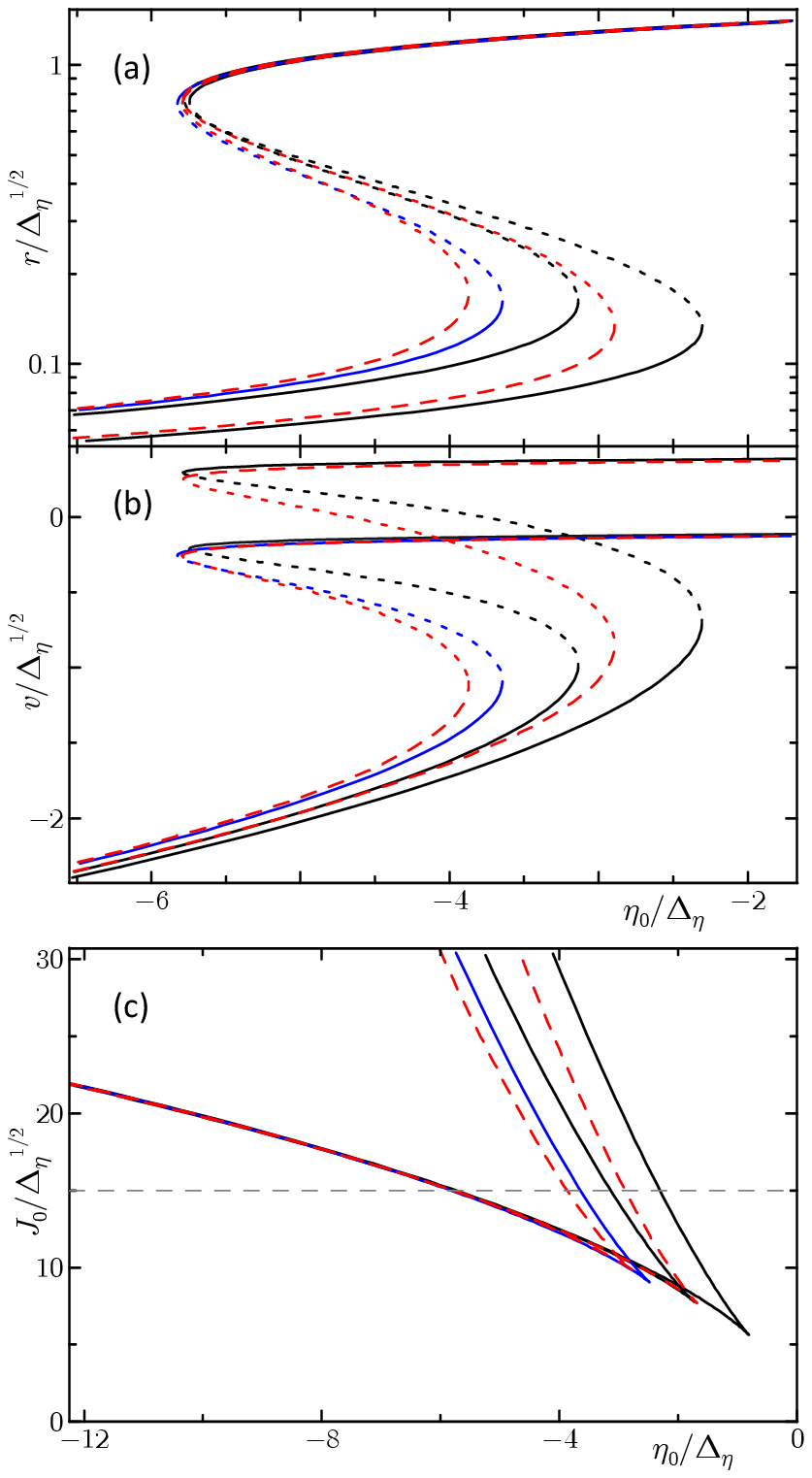}
}
\caption{Population firing rate $r$ (a) and mean potential $v$ (b) are plotted {\it versus} $\eta_0$ for a network with electrical and chemical synapses [equation system~(\ref{eq401})] with $J_0/\Delta_\eta^{1/2}=15$, $\Delta_J=0$,
$D/\Delta_\eta^{3/2}=0$ (black solid lines), $1$ (red dashed lines), electrical coupling $a/\Delta_\eta^{1/2}=0$ and $1$ (from left to right),
blue solid lines: $D/\Delta_\eta^{3/2}=a/\Delta_\eta^{1/2}=1$ and $\Delta_J/\Delta_\eta^{1/2}=\pi$;
dotted lines: the unstable branch of solution. The hysteretic transition domain is within the wedges in panel (c).
}
  \label{fig5}
\end{figure}

For $U_j=V_j-a/2$ and $u=v-a/2$, we can utilize the derived analytical solutions
to obtain a self-consistent solution for the macroscopic states of a thermodynamically large ($N\gg1$) population, for which the firing rate $r$ and potential $v$ can be treated as constants.
Given Lorentzian distributions of $\eta_j$ and $J_j$, the effective parameter $A$ remains Lorentzian, and thus the macroscopic state is described by the following system of equations:
\begin{equation}
\left\{
\begin{array}{l}
\displaystyle
\mathcal{A}_0=\frac{\eta_0+J_0r+au+a^2/4}{D^{2/3}}\,,
\quad \\[5pt]
\displaystyle
\Delta_\mathcal{A}=\frac{\Delta_\eta+\Delta_Jr}{D^{2/3}}\,,   \\[5pt]
\displaystyle
r=\frac{1}{\pi}D^{1/3}\mathrm{Re}\,\mathcal{W}\!\left(\mathcal{A}_0+i\Delta_\mathcal{A}\right),   \\[5pt]
\displaystyle
u=-D^{1/3}\mathrm{Im}\,\mathcal{W}\!\left(\mathcal{A}_0+i\Delta_\mathcal{A}\right).
\end{array}
\right.
\label{eq401}
\end{equation}
The solution to this system can be found in parametric form, parameterized by $\mathcal{A}_0$ and $\Delta_\mathcal{A}$.
Specifically, if $\Delta_J\ne0$, for a given $\mathcal{A}_0+i\Delta_\mathcal{A}$, we calculate
$$
\Delta_J=
\frac{D^{2/3}\Delta_\mathcal{A}-\Delta_\eta} {\pi^{-1}D^{1/3}\mathrm{Re}\,\mathcal{W}(\mathcal{A}_0+i\Delta_\mathcal{A})}
$$
and
$$
J_0=
\frac{D^{2/3}\mathcal{A}_0-\eta_0 +aD^{1/3}\mathrm{Im}\,\mathcal{W}(\mathcal{A}_0+i\Delta_\mathcal{A}) -a^2/4} {\pi^{-1}D^{1/3}\mathrm{Re}\,\mathcal{W}(\mathcal{A}_0+i\Delta_\mathcal{A})}
,
$$
alternatively, instead of the latter one,
$$
\eta_0=
D^{2/3}\mathcal{A}_0 -D^{1/3}\mathrm{Re}\!\left[\left(\frac{J_0}{\pi}+ia\right)\mathcal{W}(\mathcal{A}_0+i\Delta_\mathcal{A})\right] -\frac{a^2}{4}.
$$
If $\Delta_J=0$, for a given $\mathcal{A}_0$, we calculate $\Delta_\mathcal{A}=\Delta_\eta/D^{2/3}$ and use the same expression for $J_0$ (or $\eta_0$).
In Fig.\,\ref{fig5}(a,b) one can see the zone of bistability between two macroscopic regimes and hysteretic transitions.

Using solution~(\ref{eq401}), we can find the boundaries of the domain of hysteretic transitions as follows.
On the parameter plane $(\eta_0,J_0)$, at a tangential bifurcation point (solution folding), we have $\mathrm{d}\eta_0=\mathrm{d}J_0=0$ for nonzero $\mathrm{d}r\ne0$ and $\mathrm{d}u\ne0$ (hence, $\mathrm{d}\mathcal{A}_0\ne0$ and $\mathrm{d}\Delta_\mathcal{A}\ne0$).
For system~(\ref{eq401}), we have
\begin{align}
D^{2/3}\mathrm{d}\mathcal{A}_0&=J_0\mathrm{d}r+a\mathrm{d}u\,,
\nonumber\\
D^{2/3}\mathrm{d}\Delta_\mathcal{A}&=\Delta_J\mathrm{d}r\,,
\nonumber\\
\pi\mathrm{d}r-\mathrm{d}u&=D^{1/3}(\mathrm{d}\mathcal{A}_0+i\mathrm{d}\Delta_\mathcal{A})\mathcal{W}^\prime.
\nonumber
\end{align}
The solvability condition for this set of linear equations is
\begin{align}
J_0&=
\frac{
\pi D^{1/3}+(\pi a+\Delta_J)\mathrm{Im}\mathcal{W}^\prime
+\frac{a\Delta_J}{D^{1/3}}|\mathcal{W}^\prime|^2} {\mathrm{Re}\mathcal{W}^\prime}.
\label{eq402}
\end{align}
With $J_0$ given by (\ref{eq402}), the system of equations (\ref{eq401}) yields $(r,u)$ and equation for $\eta_0$:
\begin{align}
\eta_0&=D^{2/3}\mathcal{A}_0-D^{1/3}\mathrm{Re}\left[\left(\frac{J_0}{\pi}+ia\right)\mathcal{W}\right] -\frac{a^2}{4}.
\label{eq402eta}
\end{align}
From Eq.\,(\ref{eq210}), we can also derive $\mathcal{W}^\prime=i(\mathcal{A}-\mathcal{W}^2)$.
Expressions~(\ref{eq402}) and (\ref{eq402eta}) for $J_0$ and $\eta_0$,
parameterized by $\mathcal{A}_0$ and $\Delta_\mathcal{A}$ (if $\Delta_J=0$ then $\Delta_\mathcal{A}=\Delta_\eta/D^{2/3}$ is fixed), define the boundary of the hysteretic transition domain.

In the limit of vanishing noise ($D\to0$), equation system~(\ref{eq401}) is equivalent to the solution derived in Sec.\,\ref{ssec51a:MFM1a} within the framework of the MPR theory~\cite{Montbrio-Pazo-Roxin-2015,Montbrio-Pazo-2020}.

In Fig.\,\ref{fig5}(a,b), one can distinguish between the high- and low-firing-rate branches of solution. The former corresponds to $A_0>0$, which indicates the predominance of ``tonic neurons''---QIFs in a state of periodic spiking slightly perturbed by noise. In this state, the population firing rate is nearly insensitive to the noise intensity $D$, electrical coupling $a$, and synapse heterogeneity $\Delta_J$. This regime can be classified as mean-filed driven. The population-mean potential $v$ in this regime approaches $-a/2$ as $\eta_0/\Delta_\eta$ increases (see Fig.\,\ref{fig5}(b)).
The low-firing-rate regime corresponds to $A_0<0$, indicating the predominance of QIFs in the excitable state. This regime can be identified as fluctuation-driven. In Fig.\,\ref{fig5}(a), one can see that the population firing rate in this regime is increased by noise intensity $D$ and synaptic heterogeneity $\Delta_J$, while being suppressed by electrical coupling $a$.
The increase in the population firing rate can destroy this regime, causing its domain to decrease. Conversely, suppression of the population firing rate expands the domain where this regime exists, as observed in Fig.\,\ref{fig5}. The position of the tangential bifurcation point of this regime is significantly influenced by each parameter: $D$, $\Delta_J$, and $a$. Away from the bifurcation point, the population-mean potential $v$ exhibits weak but noticeable sensitivity to each of these parameter.

For $D/\Delta_\eta^{3/2}=1$, the case of $\Delta_J/\Delta_\eta^{1/2}=\pi a/\Delta_\eta^{1/2}=\pi$ (blue solid line) deviates from the case of $\Delta_J=a=0$ (the nearest red dashed line). Comparing this to the distant red dashed line ($\Delta_J=0$, $a/\Delta_\eta^{1/2}=1$), we observe that, in the presence of noise, the synaptic heterogeneity $\Delta_J$ diminishes but does not completely negate the effect of coupling $a$, as occurs when $D=0$.
The impact of electrical synapses is more complex than in the $D=0$ case, where it can be represented by recalibrating mean-field variables and parameters.

\subsection{Balanced QIF networks with sparse inhibitory synaptic links and endogenous noise}
\label{ssec52:MFM2}

In this section, we examine the implementation of mean-field theory to a more challenging case where the network dynamics are fully deterministic and quenched disorder is present in both the network geometry and unit parameters. The network irregularity and link sparseness effectively create spontaneous fluctuations in the synaptic input of individual neurons, which can be represented by an effective noise whose intensity is controlled by the instantaneous network activity~\cite{Volo-Torcini-2018,Goldobin-etal-2024,Goldobin-etal-PRE-2025}. We will employ the analytical solution for the mean-field model of this system and test it against the results of network simulations.

For a population of $N$ QIFs with a sparse network of {\em inhibitory} synaptic links, we can write 
Eq.\,(\ref{eq101}) 
with
\begin{align}
&I_j=i_0\sqrt{K_0}-\sum_{k=1}^{N}\sum_nJ_{jk}\epsilon_{jk}\delta(t-t_k^{(n)})\,,
\label{eq411}
\end{align}
where $i_0\sqrt{K_0}$: an external DC current, $J_{jk}=g_{jk}/\sqrt{K_0}$: the synaptic coupling strength, the adjacency matrix element $\epsilon_{jk}$ is $1$ if a synaptic link from the $k$-th neuron to the $j$-th one exists and $0$ otherwise; $K_j=\sum_k\epsilon_{jk}$ is the in- degree of the $j$-th neuron. The scaling of the external current and $J_{jk}$ with the mean in- degree $K_0$ is introduced explicitly in order to have a self-sustained balance between excitation and inhibition for large $K_0$~\cite{Vreeswijk-Sompolinsky-1996,Kadmon-Sompolinsky-2015}. In analogy with Erd\"os--Renyi networks, we consider the half-width of the in- degree distribution $\Delta_K=\Delta_0/\sqrt{K_0}$~\cite{Volo-Torcini-2018,Bi-Segneri-Volo-Torcini-2020}.

The irregularity of synaptic pulses effectively generates endogenous noise, for which we can adopt the diffusion approximation (DA)~\cite{Capocelli-Ricciardi-1971,Volo-etal-2022}. While caution is required when applying this approximation for small $i_0$ and $K_0\lesssim200$~\cite{Goldobin-etal-2024,Goldobin-etal-PRE-2025}, it remains accurate for
$$
\frac{K_0^3}{32}\left(\frac{i_0}{g_{jk}^2}\right)^2\frac{1}{\ln(g_{jk}^2/i_0)}\gg1
$$
in the case of macroscopically stationary states of a homogeneous population (see Eq.\,(30) in~\cite{Ageeva-Goldobin-EPJST-2026}). For a homogeneous population with $g_{jk}=g_0$, the DA effectively yields mean-field dynamics~\cite{Volo-etal-2022} with
$$
A_0=\sqrt{K_0}(i_0-g_0r) \quad \mbox{ and } \quad D=\frac{g_0^2r}{2} \, .
$$
For a heterogeneous population, where $g_{jk}$ and $i_j$ are Lorentzian-distributed around $g_0$ and $i_0$ with half-widths $\Delta_g$ and $\Delta_{i_0}$, respectively, we write
$$
A_j=i_j\sqrt{K_0}-K_j\frac{\sum_k\epsilon_{jk}g_{jk}}{K_j\sqrt{K_0}}r \, ,
\quad
D_j=\sum_k\frac{\epsilon_{jk}}{2}\frac{g_{jk}^2}{K_0}r
$$
and can derive (see Appendix~\ref{secA1}):
\begin{align}
\mathcal{A}_0+i\Delta_\mathcal{A} &=\frac{\sqrt{K_0}\big[i_0+i\Delta_{i_0} -\big(1-\frac{i\Delta_0}{\sqrt{K_0}}\big)(g_0-i\Delta_g)r\big]}{D_h^{2/3}},
\nonumber
\\
D_h &=\left(1-\frac{i\Delta_0}{\sqrt{K_0}}\right)\frac{(g_0-i\Delta_g)^2r}{2}\,,
\nonumber
\\[7pt]
\pi r-iv &=D_h^{1/3}\mathcal{W}(\mathcal{A}_0+i\Delta_\mathcal{A})\,.
\label{eq412}
\end{align}

\begin{figure}[!t]
\centerline{
\includegraphics[width=0.995\columnwidth]%
 {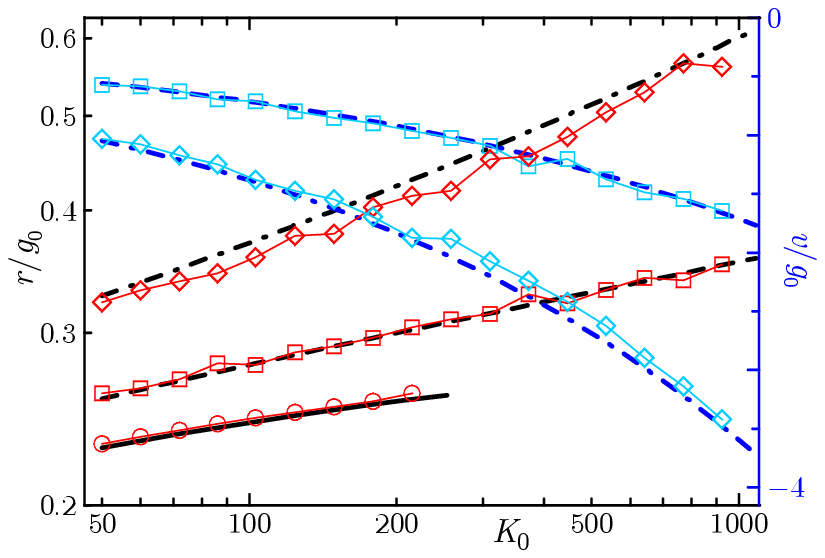}
}
\caption{For a sparse balanced network of neurons~(\ref{eq411}), analytical mean-field solutions~(\ref{eq412}) (lines) are compared to the results of direct numerical simulations of the network of $N=10\,000$ neurons (symbols) for $i_0/g_0=0.3$. Circles and solid line: firing rate of a homogeneous population; squares, dashed lines: firing rate $r$ (red, black) and population-mean membrane potential $v$ (blue) for a population with heterogeneity of DC current $\Delta_{i_0}=0.4i_0$, $\Delta_0=\Delta_g=0$; diamonds, dash-dotted lines: $r$ (red, black) and $v$ (blue) for $\Delta_{i_0}=0.4i_0$ and heterogeneities of in- degrees and synaptic link strengths, $\Delta_0=0.5$, $\Delta_g/g_0=0.5$.
For a homogeneous network, the curve and symbols are plotted only for $K_0\lesssim250$, since the time-independent solution becomes unstable and collective oscillations occur for higher $K_0$~\cite{Goldobin-etal-2024,Goldobin-etal-PRE-2025}.
}
  \label{fig6}
\end{figure}

In Fig.\,\ref{fig6}, one can assess the agreement between this solution and the results of finite network simulations.
For the network simulations with heterogeneous in- degree $K_j$, we first generate the value of in- degree of the node $j$ from a Lorentzian distribution around $K_0$ with half-width $\Delta_0\sqrt{K_0}$; second, we take the nearest integer; third, negative $K_j$ and (for symmetry) $K_j>2K_0$ are discarded and regenerated. The network simulation results are averaged over 10 realizations of the synaptic network.

There are three primary sources of discrepancy between the network simulations and the mean-field theory:
\begin{enumerate}[label=(\roman*)]
\item
\label{list:1}
the inaccuracy of the diffusion approximation for noise, which is a Poissonian shot noise~\cite{Goldobin-etal-2024,Goldobin-etal-PRE-2025};
\item
\label{list:2}
finite-size effects in the network ($N=10^4$);
\item
\label{list:3}
discreteness of in- degrees $K_j$ (rounding to the nearest integer) and cut-off of values $K_j<0$ and $K_j>2K_0$ given by a Lorentzian distribution.
\end{enumerate}
Regarding \ref{list:2}, the condition for the validity of the Poissonian noise approximation is the absence of correlations between synaptic signals received from different neurons. This requires $K/N\ll1$~\cite{Goldobin-etal-2024,Goldobin-etal-PRE-2025}. In this respect, the network size $N=10^4$ is actually quite far from an idealistic thermodynamic limit of a large population, especially for large in- degree: for $K_0$ in Fig.\,\ref{fig6}, the ratio $K_0/N$ ranges from $0.005$ to $0.1$.
The results for $\Delta_0=0$ are affected only by the first two sources of discrepancy. The magnitude of these effects can be assessed by comparing the solid (dashed) lines with the circles (squares) in Fig.\,\ref{fig6}. Source~\ref{list:3} affects the case of $\Delta_0=0.5$ and its impact can be observed in the reduced agreement between the dash-dotted lines and diamond symbols compared to the cases with $\Delta_0=0$.
Overall, despite these discrepancies, the mean-field theory provides a reasonable representation of the finite network behavior as shown in Fig.\,\ref{fig6}.

The reasonably good agreement between the analytical mean-field theory and the results of direct numerical network simulations observed in Fig.\,\ref{fig6} serves as an important validation of the theoretical approach. Here, the validity of the mean-field theory is provided by the laws of large numbers, while these numbers are quite modestly large. The switching to the mean-field theory is conducted in two steps:
(i)
the irregularity of the network geometry and incoming synaptic activity are represented by an effective noise;
(ii)
the distributed effective parameters $A_j$ and effective noise intensity $D_j$ are introduced.
The distribution of effective $A_j$ and $D_j$ is derived theoretically, not by fitting to the simulation results.
Moreover, the Residue theorem cannot be directly applied to the diffusion-approximation model with Lorentzian-distributed intensity $D_j$, but it can be applied to the shot-noise model, with the diffusion approximation introduced subsequently, as detailed in Appendix~\ref{secA1}. The results of network simulations in Fig.\,\ref{fig6} confirm the validity of this approach.


\section{Conclusion}\label{sec6:Concl}
We have derived an analytical solution for the first passage time problem for a quadratic integrate-and-fire neuron (QIF) subject to white Gaussian noise. This solution provides an expression for the mean interspike interval, which can be further used to calculate the firing rate $r$ and population-mean membrane potential $v$ in a homogeneous population. The analytic form of the Montbri\'o--Paz\'o--Roxin parameter~\cite{Montbrio-Pazo-Roxin-2015} (which is also the first pseudocumulant~\cite{Goldobin-Volo-Torcini-2021}) $W=\pi r-iv=D^{1/3}\mathcal{W}(A/D^{2/3})$ (\ref{eqW2}) allows generalization of the results to heterogeneous populations with Lorentzian and other~\cite{Klinshov-Kirillov-Nekorkin-2021} distributions of parameters [see Eq.\,(\ref{eq303}) and Fig.\,\ref{fig3}].

Using these analytical results, we studied the effect of noise on the interplay between electrical and chemical synapses in QIF networks. The mean-field theory solution is given by (\ref{eq401}); the boundaries of the domain of hysteretic transitions between the fluctuation- and mean-field-driven regimes are given by Eqs.\,(\ref{eq402}) and (\ref{eq402eta}) (Fig.\,\ref{fig5}).
For the case of sparse inhibitory networks, which require a significantly more sophisticated mean-field theoretical treatment, the applicability of the effective noise model and the corresponding mean-field theory for networks with parameter heterogeneities and node in- degree $K$ heterogeneity has been validated trough finite network simulations.

The presented solutions for time-independent macroscopic regimes are also applicable for quasi-stationary (adiabatically changing) network configurations. For instance, they can be practically useful for theoretical studies of neural plasticity~\cite{Tsodyks-Markram-1997,Bi-Poo-1998} in QIF networks and slow adaptation in systems of coupled rotators or excitable elements~\cite{Sawicki-etal-adapt-2023,Ratas-Pyragas-2016,Taher-Torcini-Olmi-2020,Gast-Knosche-Schmidt-2021, Franovic-etal-2022,Eydam-Franovic-Kang-2024,Klinshov-Nekorkin-2024,Bergoin-Torcini-etal-2025}.

The statistical theory problem of QIFs is mathematically equivalent to the problem of Anderson localization in the one-dimensional Schr\"odinger equation with a frozen random potential (quenched disorder)~\cite{Frisch-Lloyd-1960,Halperin-1965,Lifshitz-Gredeskul-Pastur-1988} (Appendix~\ref{secA2}).
For the one-dimensional Ising model, the problem of calculating the partition function and the specific free energy is equivalent to calculating the Lyapunov exponent~\cite{Derrida-Hilhorst-1983,Mello-Robledo-1993,Zillmer-Pikovsky-2005}.

%

\begin{acknowledgments}
We are thankful to A.\ Pikovsky, E.\,V.\ Shklyaeva, A.\,V.\ Dolmatova, and L.\,S.\ Klimenko for discussions that made this work possible.
The theoretical study presented in Secs.~\ref{sec2:QIF}--\ref{sec4:Heter}
was financially supported by the Russian Science Foundation (Grant No.\ 23-12-00180-P).
\end{acknowledgments}

\appendix
\section{Averaging over Lorentzian-distributed parameters for a shot noise}
\label{secA1}
For Eq.\,(\ref{eq101}) with current~(\ref{eq411}), approximated by Poissonian trains of $\delta$-pulses (shot noise),
the continuity equation for PDF $P_j(V_j,t)$~\cite{Gardiner-1997,Goldobin-etal-2024,Goldobin-etal-PRE-2025,Ageeva-Goldobin-EPJST-2026} reads
\begin{align}
\frac{\partial P_j(V_j,t)}{\partial t} &+\frac{\partial}{\partial V_j}\big[(A_j^{(c)}+V_j^2)P_j(V_j,t)\big]
\nonumber\\
&=\sum_k\epsilon_{jk}r_k(t)\big[P_j(V_j+J_{jk},t)-P_j(V_j,t)\big]\,,
\nonumber
\end{align}
where $A_j^{(c)}$ stands for the constant part of the current. For $N\gg1$, the network of input synaptic links is effectively averaged over the entire population and one can replace $r_k(t)$ with the population-mean $r(t)$, and rewrite the continuity equation in Fourier space;
\begin{align}
\frac{\partial\mathcal{P}_j(q,t)}{\partial t} &-iq\int_{-\infty}^{+\infty}\!\!e^{iqV_j}(A_j^{(c)}+V_j^2)P_j(V_j,t)\,\mathrm{d}q
\nonumber\\
&=r(t)\sum_k\epsilon_{jk}\left(e^{-iqJ_{jk}}-1\right)\mathcal{P}_j(q,t)\,,
\label{eqSMA1}
\end{align}
where $P_j(V_j,t)=(2\pi)^{-1}\int_{-\infty}^{+\infty}\mathcal{P}_j(q,t)e^{-iqV_j}\mathrm{d}q$ [we use this definition to have the characteristic function of variable $V_j$: $\mathcal{P}_j(q,t)=\int_{-\infty}^{+\infty}P_j(V_j,t)\,e^{iqV_j}\mathrm{d}V_j$]. The shot-noise term can be averaged over Lorentzian-distributed $J_{jk}$ with closure of the integration contour over the lower arc on the complex plane of $J$ for $q>0$ since $e^{-iqJ}\to0$ for $J\to-i\infty$, and over the upper one otherwise; hence,
$$
\sum_k\epsilon_{jk}(e^{-iqJ_{jk}}-1)\mathcal{P}_j(q,t) \to K_j(e^{-iqJ_0-\Delta_J|q|}-1)\mathcal{P}_j(q,t).
$$
If we could limit our consideration to $q>0$, then we would have the noise term with $J_{jk}\mapsto J_0-i\Delta_J$ after averaging.
Further, averaging over Lorentzian-distributed $K_j$ is trivial, since the real part of $(e^{-iqJ_0-\Delta_J|q|}-1)$ is nonpositive and the imaginary part is nonpositive for $qJ_0\in[0;\pi]$; with $q>0$ and small $qJ_0=qg_0/\sqrt{K_0}\le\pi$ we should close over the lower half-plane of complex $K$: $K_j\mapsto K_0-i\Delta_0\sqrt{K_0}$.

To confirm that, in our specific problem, we can use the results for $q>0$, we need to switch to the phase variable $\theta_j=2\arctan{V_j}$. In the Fourier space of PDF of $\theta_j$, the dynamics of $Z_n=\langle{e^{in\theta}}\rangle$ for $n>0$ ($Z_0=1$ is constant) is contributed only by $Z_n$ and $Z_{n\pm1}$ for the deterministic part (as in the Ott--Antonsen theory~\cite{Ott-Antonsen-2008}), and only by $Z_m$ with $m>0$ for the shot-noise terms (for instance, see~\cite{Goldobin-etal-2024,Goldobin-etal-PRE-2025}). Hence, the mathematical description can be confined to $Z_n$ with $n>0$, for which the consideration of the closure of integrals over the Lorentzian-distributed parameters for the shot-noise terms is much more involved but geometrically the same as for Eq.\,(\ref{eqSMA1}) with $q>0$. Simultaneously, the MPR order parameter $W=1-2Z_1+2Z_2-2Z_3+\cdots$~\cite{Montbrio-Pazo-Roxin-2015} is calculated with $Z_{n>0}$ only. Thus, the averaging of $W$ over Lorentzian-distributed $J_{jk}$ and $K_j$ can be done and it involves poles $(g_0-i\Delta_g)/\sqrt{K_0}$ and $K_0-i\Delta_0\sqrt{K_0}$.

With these results we can introduce the DA for Eq.\,(\ref{eqSMA1}), which amounts to the Taylor expansion of the noise term with respect to $q$ and truncation after the second-derivative term. Hence, the first-derivative term contributes to $A_j$,
$$
A_{j}^+=-r(\sqrt{K_0}-i\Delta_0)(g_0-i\Delta_g)
$$
and the second-derivative term gives effective complex
$$
D=\left(1-\frac{i\Delta_0}{\sqrt{K_0}}\right)\left(g_0-i\Delta_g\right)^2\frac{r}{2} \, .
$$

Notice, similar averaging cannot be performed starting from the diffusion-approximation equations, since it formally requires dealing with negative values of the diffusion coefficient $D$, which are senseless physically and lead to infinitely fast divergence mathematically. One must first conduct the averaging for the first-principle model ({\it e.g.}, shot-noise one) and only then use the DA for the results of the first step.

\section{Anderson localization in one-dimensional Schr\"odinger equation}
\label{secA2}
In this section we explain the connection between the studied problem and its quantum mechanics counterpart. We show that the localization properties of states in the 1D Schr\"odinger equation (SE) with a frozen random potential are governed by a mathematical problem which is equivalent to the problem of QIFs.
The dimensionless Schr\"odinger equation for a particle in a frozen 1D potential landscape $\sqrt{D}\zeta(x)$, $\langle\zeta(x)\,\zeta(x+x^\prime)\rangle=2\delta(x^\prime)$~\cite{Frisch-Lloyd-1960,Halperin-1965}, reads
\begin{equation}
-i\frac{\partial}{\partial t}\psi(x,t)=-\frac{\partial^2}{\partial x^2}\psi(x,t)+\sqrt{D}\zeta(x)\psi(x,t)\,.
\label{eq:a01}
\end{equation}
For a state with energy $E$, the wave function $\psi(x,t)=e^{iEt}\psi_E(x)$ and $\psi_E(x)$ obeys the stationary Schr\"odinger equation:
\begin{equation}
\psi_E^{\prime\prime}(x)=-E\psi_E(x)+\sqrt{D}\zeta(x)\psi_E(x)\,.
\label{eq:a02}
\end{equation}
For the logarithmic derivative $z=\psi_E^\prime/\psi_E$, Eq.\,(\ref{eq:a02}) yields \cite{Lifshitz-Gredeskul-Pastur-1988,Zillmer-Pikovsky-2003,Pikovsky-Politi-LE_book-2016}
\begin{equation}
z^\prime=-z^2-E+\sqrt{D}\zeta(x)\,.
\label{eq:a03}
\end{equation}
This equation is equivalent to Eqs.\,(\ref{eq101})--(\ref{eq102}), up to the substitution \begin{equation}
(z,E,x)\to(-V,A,t)\,.
\label{eq:a04}
\end{equation}

Let us now interpret ordinary differential equation~(\ref{eq:a03}) in context of spatial problems (\ref{eq:a01}) and (\ref{eq:a02}) with boundary conditions. By virtue of the statistical symmetry $(x\leftrightarrow-x)$ of Eqs.\,(\ref{eq:a01}) and (\ref{eq:a02}), the modes $\psi_{E,\pm}(x)=\Psi_{E,\pm}(x)e^{\pm\lambda_E x}$, where $\lambda_E>0$ and $\Psi_{E,\pm}(x)$ neither grow nor decay on average over long distances, emerge always in pairs. States in (\ref{eq:a01}) or (\ref{eq:a02}) on an infinite line should be bounded at infinity; therefore, under sensible boundary conditions for (\ref{eq:a01}) or (\ref{eq:a02}), one can observe only the $\Psi_{E,+}(x)e^{\lambda_E x}$ mode for $x\to-\infty$ and the $\Psi_{E,-}(x)e^{-\lambda_E x}$ mode for $x\to+\infty$. These two tails are combined for the state of energy $E$: $\psi_E(x)=\Psi_E(x)e^{-\lambda_E|x-x_E|}$, where $x_E$ is the localization center of the state and $\lambda_E$ is a {\em localization exponent}. Simultaneously, for the ordinary differential equation~(\ref{eq:a02}) with some initial conditions at $x_0$, the mode $\psi_{E,+}(x>x_0)$ is growing, while $\psi_{E,-}(x>x_0)$ is decaying. Therefore, on average over long intervals of $x$, one will observe the $\psi_{E,+}$ mode; for this mode
\[
\overline{\frac{\mathrm{d}}{\mathrm{d}x}\ln\psi}=\overline{z}=\lambda_E\,,
\]
where $\overline{(\cdot)}$ indicates averaging along $x$. With such a formulation, $\lambda_E$ can be interpreted as the Lyapunov exponent~\cite{Zillmer-Pikovsky-2003,Pikovsky-Politi-LE_book-2016}. Due to ergodicity of the problem of calculation of $\overline{z}$ along $x$ in Eq.\,(\ref{eq:a03}), the $x$-average is equivalent to the average over noise realizations:
\begin{equation}
\lambda_E=\langle{z}\rangle\,.
\label{eq:a05}
\end{equation}

Further, in the noise-free case, the solution of Eq.\,(\ref{eq:a02}) is $\cos{k_E(x-x_E)}$ with $k_E=\sqrt{E}$. In the noisy case the solution envelope will decay at distance owned by localization, but local oscillations will be still present, $\Psi_E(x)\sim\cos{\phi_E(x)}$, and $\Psi_E(x)$ will transpass $0$ twice per one turnover period. As $\psi_E(x)$ transpasses $0$, $\psi_E(x)\approx\psi_E^\prime(x_0)(x-x_0)$ and $z\approx1/(x-x_0)$, {\it i.e.}, $z$ is reset from $-\infty$ to $+\infty$ twice per the wave function turnover period. This is equivalent to two firing events of a QIF and the reference wavenumber of the mode
\begin{equation}
\frac{k_{E,\ast}}{2\pi}=\frac{r}{2}\,.
\label{eq:a06}
\end{equation}
Moreover, Frisch and Lloyd~\cite{Frisch-Lloyd-1960} derived that the cumulative density of states $N(E)\equiv\int_{-\infty}^E\rho(E_1)\mathrm{d}E_1$ (where $\rho(E)$ is the density of the energy levels) is given by the limit $\lim_{z\to\pm\infty}z^2P(z)$, which is $r$ in the case of QIFs.

Comparing Eq.\,(\ref{eq:a03}) to Eqs.\,(\ref{eq101})--(\ref{eq102}), one can make use of transformation~(\ref{eq:a04}) in order to employ results~(\ref{eq212}) and (\ref{eq213}) for Eqs.\,(\ref{eq:a05}) and (\ref{eq:a06}). One obtains
\begin{align}
&
 \pi N(E) + i\lambda_E=D^{1/3}\mathcal{W}\big(E/D^{2/3}\big)\,,
\label{eq:a07}\\
&\mathcal{W}(x)=
\frac{\sqrt{x}}{i}
\frac{\mathrm{J}_{2/3}(\frac{2}{3}e^{i\pi}x^{3/2}) -\mathrm{J}_{-2/3}(\frac{2}{3}e^{i\pi}x^{3/2})}
{\mathrm{J}_{1/3}(\frac{2}{3}e^{i\pi}x^{3/2}) +\mathrm{J}_{-1/3}(\frac{2}{3}e^{i\pi}x^{3/2})}\,.
\label{eq:a08}
\end{align}

Here we have explicitly shown how the problem of statistical theory of QIFs is mathematically identical to the problem of Anderson localization in SE with quenched disorder. For SE, we recall that the theory also delivers the cumulative density of the energetic levels.
It is also noteworthy, that for the one-dimensional Ising model, the problem of calculation of the partition function and the specific free energy is equivalent to the problem of calculation of the Lyapunov exponent~\cite{Derrida-Hilhorst-1983,Mello-Robledo-1993,Zillmer-Pikovsky-2005}.


\begin{thebibliography}{20}

\bibitem{Ermentrout-Kopell-1986}
 G.~B.\ Ermentrout and N.\ Kopell,
 {Parabolic bursting in an excitable system coupled with a slow oscillation},
 SIAM J.\ Appl.\ Math. {\bf 46}, 233 
 (1986).
 \\
 \url{https://doi.org/10.1137/0146017}

\bibitem{Izhikevich-2007}
 E.~M.\ Izhikevich,
 {\em Dynamical systems in neuroscience}
 (MIT Press, Cambridge, MA, 2007).



\bibitem{Pikovsky-Politi-LE_book-2016}
 A.\ Pikovsky and A.\ Politi,
 {\em Lyapunov Exponents: A Tool to Explore Complex Dynamics}
 (Cambridge University Press, Cambridge, 2016).
 \\
 \url{https://doi.org/10.1017/CBO9781139343473}


\bibitem{Klyatskin-2017}
 V.~I.\ Klyatskin,
 {Stochastic parametric resonance}, in
 \textit{Fundamentals of Stochastic Nature Sciences}
 (Springer, Cham, 2017) pp.\ 55--57.\\
 \url{https://doi.org/10.1007/978-3-319-56922-2_6}

\bibitem{Klyatskin-1975}
 V.~I.\ Klyatskin,
 \textit{Statisticheskoe opisanie sistem s fluktuiruyushchimi parametrami}
 [in Russian; English translation: \textit{Statistical Description of Dynamic Systems with Fluctuating Parameters}]
 (Nauka, Moscow, 1975). 

\bibitem{Khasminskii-2012}
 R.\ Khasminskii,
 {\em Stochastic Stability of Differential Equations.}
 \textit{Stochastic Modelling and Applied Probability},
 2nd ed.\
 (Springer-Verlag, Berlin, Heidelberg, 2012), Vol.\ 66.
 \url{https://doi.org/10.1007/978-3-642-23280-0}

\bibitem{Zillmer-Pikovsky-2003}
 R.\ Zillmer and A.\ Pikovsky,
 {Multiscaling of noise-induced parametric instability},
 Phys.\ Rev.\ E {\bf 67}, 061117 (2003).
 \url{https://doi.org/10.1103/PhysRevE.67.061117}


\bibitem{Goldobin-etal-2017}
 D.~S.\ Goldobin, A.~V.\ Pimenova, M.\ Rosenblum, and A.\ Pikovsky,
 {Competing influence of common noise and desynchronizing coupling on synchronization in the Kuramoto-Sakaguchi ensemble},
 Eur.\ Phys.\ J.\ Spec.\ Top. {\bf 226}, 1921 
 (2017).
 \\
 \url{https://doi.org/10.1140/epjst/e2017-70039-y}

\bibitem{Permyakova-Goldobin-2023}
 E.~V.\ Permyakova and D.~S.\ Goldobin,
 {Stochastic parametric excitation of convective heat transfer},
 Phil.\ Trans.\ R.\ Soc.\ A {\bf 381}, 20220084 (2023).\\
 \url{https://doi.org/10.1098/rsta.2022.0084}



\bibitem{Lifshitz-Gredeskul-Pastur-1988}
 I.~M.\ Lifshitz, S.~A.\ Gredeskul, and L.~A.\ Pastur,
 {\em Introduction to the Theory of Disordered Systems}
 (Wiley, New York, 1988).

\bibitem{Frisch-Lloyd-1960}
 H.~L.\ Frisch and S.~P.\ Lloyd,
 {Electron levels in a one-di\-men\-sio\-nal random lattice},
 Phys.\ Rev.\ {\bf 120}, 1175 
 (1960).
 \\
 \url{https://doi.org/10.1103/PhysRev.120.1175}

\bibitem{Halperin-1965}
 B.~I.\ Halperin,
 {Green's functions for a particle in a one-di\-mensional random potential},
 Phys.\ Rev. {\bf 139}, A104 
 (1965).
 \url{https://doi.org/10.1103/PhysRev.139.A104}


\bibitem{Zillmer-Pikovsky-2005}
 R.\ Zillmer and A.\ Pikovsky,
 {Continuous approach for the random-field Ising chain},
 Phys.\ Rev.\ E {\bf 72}, 056108 (2005).
\\
 \url{https://doi.org/10.1103/PhysRevE.72.056108}

\bibitem{Derrida-Hilhorst-1983}
 B.\ Derrida and H.~J.\ Hilhorst,
 {Singular behaviour of certain infinite products of random 2 {\rm x} 2 matrices},
 J.\ Phys.\ A {\bf 16}, 2641 
 (1983).\\
 \url{https://doi.org/10.1088/0305-4470/16/12/013}

\bibitem{Mello-Robledo-1993}
 P.~A.\ Mello and A.\ Robledo,
 {Strongly coupled Ising chain under a weak random field},
 Physica A {\bf 199}, 
 363 
 (1993).
 \url{https://doi.org/10.1016/0378-4371(93)90055-9}


\bibitem{Goldobin-Shklyaeva-2013}
 D.~S.\ Goldobin and E.~V.\ Shklyaeva,
 {Localization and advectional spreading of convective flows under parametric disorder},
 J.\ Stat.\ Mech. P09027 (2013).\\
 \url{https://doi.org/10.1088/1742-5468/2013/09/P09027}

\bibitem{Goldobin-Shklyaeva-2009}
 D.~S.\ Goldobin and E.~V.\ Shklyaeva,
 {Diffusion of a passive scalar by convective flows under parametric disorder},
 J.\ Stat.\ Mech. P01024 (2009).\\
 \url{https://doi.org/10.1088/1742-5468/2009/01/P01024}

\bibitem{Goldobin-2010}
 D.~S.\ Goldobin,
 {Advectional enhancement of eddy diffusivity under parametric disorder},
 Phys.\ Scr.\ {\bf T142}, 014050 (2010).
 \url{https://doi.org/10.1088/0031-8949/2010/T142/014050}



\bibitem{Capocelli-Ricciardi-1971}
 R.\ Capocelli and L.\ Ricciardi,
 {Diffusion approximation and first passage time problem for a model neuron},
 Kybernetik {\bf 8}, 214 
 (1971).
 \\
 \url{https://doi.org/10.1007/BF00288750}


\bibitem{Ageeva-Goldobin-PREL-2026}
 A.~V.\ Ageeva and D.~S.\ Goldobin,
 {Analytical solution for mean-field theories for homo- and heterogeneous
populations of parabolic bursting systems},
 Phys.\ Rev.\ E {\bf 114}, L042105 (2026).
 \url{https://doi.org/10.1103/nl4b-7sfq}



\bibitem{Hirsch-Huberman-Scalapino-1982}
 J.~E.\ Hirsch, B.~A.\ Huberman, and D.~J.\ Scalapino,
 {Theory of intermittency},
 Phys.\ Rev.\ A {\bf 25}, 519 
 (1982).
 \\
 \url{https://doi.org/10.1103/PhysRevA.25.519}

\bibitem{Lindner-Longtin-Bulsara-2003}
 B.\ Lindner, A.\ Longtin, and A.\ Bulsara,
 {Analytic expressions for rate and CV of a type I neuron driven by white Gaussian noise},
 Neural Comput.\ {\bf 15}, 1761 
 (2003).
 \\
 \url{https://doi.org/10.1162/08997660360675035}

\bibitem{Volo-etal-2022}
 M.\ di Volo, M.\ Segneri, D.~S.\ Goldobin, A.\ Politi, and A.\ Torcini,
 {Coherent oscillations in balanced neural networks driven by endogenous fluctuations},
 Chaos {\bf 32}, 023120 (2022).
 \url{https://doi.org/10.1063/5.0075751}



\bibitem{Montbrio-Pazo-Roxin-2015}
 E.\ Montbri\'o, D.\ Paz\'o, and A.\ Roxin,
 {Macroscopic description for networks of spiking neurons},
 Phys.\ Rev.\ X {\bf 5}, 021028 (2015).\\
 \url{https://doi.org/10.1103/PhysRevX.5.021028}

\bibitem{Goldobin-Volo-Torcini-2021}
 D.~S.\ Goldobin, M.\ di Volo, and A.\ Torcini,
 {Reduction methodology for fluctuation driven population dynamics},
 Phys.\ Rev.\ Lett. {\bf 127}, 038301 (2021).\\
 \url{https://doi.org/10.1103/PhysRevLett.127.038301}



\bibitem{Gardiner-1997}
 C.~W.\ Gardiner,
 {\em Handbook of Stochastic Methods}, 2nd ed.\
 (Springer, Berlin, 1997).

\bibitem{Goldobin-2024}
 D.~S.\ Goldobin,
 {Interspike interval statistics for quadratic integrate-and-fire neurons subject to alpha-stable noise},
 Cybern.\ Phys. {\bf 13}, 206 
 (2024).
 \url{https://doi.org/10.35470/2226-4116-2024-13-3-206-210}


\bibitem{Abramowitz}
 {\em Handbook of Mathematical Functions},
 edited by M.\ Abramowitz and I.~A.\ Stegun
 (Dover, New York, 1965).




%



\bibitem{Yakubovich-1969}
 E.~I.\ Yakubovich,
 {Dynamics of processes in media with inhomogeneous broadening of the line of the working transition},
 Sov.\ Phys.\ JETP {\bf 28}, 160 
 (1969).

\bibitem{Rabinovich-Trubetskov-1989}
 M.~I.\ Rabinovich and D.~I.\ Trubetskov,
 {\em Oscillations and Waves: In Linear and Nonlinear Systems}
 (Springer Netherlands, Dordrecht, 1989), Vol.\ 50.

\bibitem{Crawford-1994}
 J.~D.\ Crawford,
 {Amplitude expansions for instabilities in populations of globally-coupled oscillators},
 J.\ Stat.\ Phys. {\bf 74}, 
 1047 
 (1994).
 \url{https://doi.org/10.1007/BF02188217}

\bibitem{Ott-Antonsen-2008}
 E.\ Ott and T.~M.\ Antonsen,
 {Low dimensional behavior of large systems of globally coupled oscillators},
 Chaos {\bf 18}, 037113 (2008).
 \url{https://doi.org/10.1063/1.2930766}


\bibitem{Klinshov-Kirillov-Nekorkin-2021}
 V.\ Klinshov, S.\ Kirillov, and V.\ Nekorkin,
 {Reduction of the collective dynamics of neural populations with realistic forms of heterogeneity},
 Phys.\ Rev.\ E {\bf 103}, L040302 (2021).
 \\
 \url{https://doi.org/10.1103/PhysRevE.103.L040302}




\bibitem{Montbrio-Pazo-2020}
 E.\ Montbri\'o and D.\ Paz\'o,
 {Exact mean-field theory explains the dual role of electrical synapses in collective synchronization},
 Phys.\ Rev.\ Lett. {\bf 125}, 248101 (2020).
 \url{https://doi.org/10.1103/PhysRevLett.125.248101}


\bibitem{Clusella-Montbrio-2024}
 P.\ Clusella and E.\ Montbri\'o,
 {Exact low-dimensional description for fast neural oscillations with low firing rates},
 Phys.\ Rev.\ E {\bf 109}, 014229 (2024).
 \\
 \url{https://doi.org/10.1103/PhysRevE.109.014229}



\bibitem{Volo-Torcini-2018}
 M.\ di~Volo and A.\ Torcini,
 {Transition from asynchronous to oscillatory dynamics in balanced spiking networks with instantaneous synapses},
 Phys.\ Rev.\ Lett. {\bf 121}, 128301 (2018).
 \\
 \url{https://doi.org/10.1103/PhysRevLett.121.128301}

\bibitem{Goldobin-etal-2024}
 D.~S.\ Goldobin, M.\ di~Volo, and A.\ Torcini,
 {Discrete synaptic events induce global oscillations in balanced neural networks},
 Phys.\ Rev.\ Lett. {\bf 133}, 238401 (2024).
 \\
 \url{https://doi.org/10.1103/PhysRevLett.133.238401}

\bibitem{Goldobin-etal-PRE-2025}
 D.~S.\ Goldobin, M.~V. Ageeva, M.\ di~Volo, F.\ Tixidre, and A.\ Torcini,
 {Synaptic shot-noise triggers fast and slow global oscillations in balanced neural networks},
 Phys.\ Rev.\ E {\bf 112}, 034301 (2025).
 \\
 \url{https://doi.org/10.1103/47h5-fbyy}


\bibitem{Vreeswijk-Sompolinsky-1996}
 C.\ van Vreeswijk and H.\ Sompolinsky,
 {Chaos in neuronal networks with balanced excitatory and inhibitory activity},
 Science {\bf 274}, 1724 (1996).\\
 \url{https://doi.org/10.1126/science.274.5293.1724}

\bibitem{Kadmon-Sompolinsky-2015}
 J.\ Kadmon and H.\ Sompolinsky,
 {Transition to chaos in random neuronal networks},
 Phys.\ Rev.\ X {\bf 5}, 041030 (2015).
 \url{https://doi.org/10.1103/PhysRevX.5.041030}


\bibitem{Bi-Segneri-Volo-Torcini-2020}
 H.\ Bi, M.\ Segneri, M.\ di Volo, and A.\ Torcini,
 {Coexistence of fast and slow gamma oscillations in one population of inhibitory spiking neurons},
 Phys.\ Rev.\ Research {\bf 2}, 013042 (2020).
 \url{https://doi.org/10.1103/PhysRevResearch.2.013042}


\bibitem{Ageeva-Goldobin-EPJST-2026}
 M.~V.\ Ageeva and D.~S.\ Goldobin,
 {Low-dimensional dynamics of a sparse balanced synaptic network of quadratic integrate-and-fire neurons},
 Eur.\ Phys.\ J.\ Spec.\ Top. {\bf 235}, 2511 
 (2026).
 \\
 \url{https://doi.org/10.1140/epjs/s11734-025-02096-8}


\bibitem{Tsodyks-Markram-1997}
 M.~V.\ Tsodyks and H.\ Markram,
 {The neural code between neocortical pyramidal neurons depends on neurotransmitter release probability},
 Proc.\ Natl.\ Acad.\ Sci.\ U.S.A. {\bf 94}, 719 
 (1997).
\\
 \url{https://doi.org/10.1073/pnas.94.2.719}


\bibitem{Bi-Poo-1998}
 G.-Q.\ Bi and M.-M.\ Poo,
 {Synaptic modifications in cultured hippocampal neurons: Dependence on spike timing, synaptic strength, and postsynaptic cell type},
 J.\ Neurosci. {\bf 18}, 10464 
 (1998).
 \url{https://doi.org/10.1523/JNEUROSCI.18-24-10464.1998}


\bibitem{Sawicki-etal-adapt-2023}
 J.\ Sawicki, R.\ Berner, S.~A.~M.\ Loos, M.\ Anvari, {\it et al.}, 
 {Perspectives on adaptive dynamical systems},
 Chaos {\bf 33}, 071501 (2023).
 \url{https://doi.org/10.1063/5.0147231}


\bibitem{Ratas-Pyragas-2016}
 I.\ Ratas and K.\ Pyragas,
 {Macroscopic self-oscillations and aging transition in a network of synaptically coupled quadratic integrate-and-fire neurons},
 Phys.\ Rev.\ E {\bf 94}, 032215 (2016).
 \\
 \url{https://doi.org/10.1103/PhysRevE.94.032215}


\bibitem{Taher-Torcini-Olmi-2020}
 H.\ Taher, A.\ Torcini, and S.\ Olmi,
 {Exact neural mass model for synaptic-based working memory},
 PLoS Comput.\ Biol. {\bf 16}, e1008533 (2020).
 \\
 \url{https://doi.org/10.1371/journal.pcbi.1008533}


\bibitem{Gast-Knosche-Schmidt-2021}
 R.\ Gast, T.~R.\ Kn\"osche, and H.\ Schmidt,
 {Mean-field approximations of networks of spiking neurons with short-term synaptic plasticity},
 Phys.\ Rev.\ E {\bf 104}, 044310 (2021).
 \\
 \url{https://doi.org/10.1103/PhysRevE.104.044310}


\bibitem{Franovic-etal-2022}
 I.\ Franovi\'c, S.\ Eydam, S.\ Yanchuk, and R.\ Berner,
 {Collective Activity Bursting in a Population of Excitable Units Adaptively Coupled to a Pool of Resources},
 Front.\ Netw.\ Physiol. {\bf 2}, 841829 (2022).
\\
 \url{doi: 10.3389/fnetp.2022.841829}


\bibitem{Eydam-Franovic-Kang-2024}
 S.\ Eydam, I.\ Franovi\'c, and L.\ Kang,
 {Control of seizure-like dynamics in neuronal populations with excitability adaptation related to ketogenic diet},
 Chaos {\bf 34}, 053128 (2024).
 \url{https://doi.org/10.1063/5.0180954}


\bibitem{Klinshov-Nekorkin-2024}
 V.~V.\ Klinshov and V.~I.\ Nekorkin,
 {Adaptive myelination causes slow oscillations in recurrent neural loops},
 Chaos {\bf 34}, 033101 (2024).
 \\
 \url{https://doi.org/10.1063/5.0193265}


\bibitem{Bergoin-Torcini-etal-2025}
 R.\ Bergoin, A.\ Torcini, G.\ Deco, M.\ Quoy, and G.\ Zamora-L\'opez,
 {Emergence and maintenance of modularity in neural networks with Hebbian and anti-Hebbian inhibitory STDP}, PLoS Comput.\,Biol.\,{\bf 21},\,e1012973\,(2025).
 \\
 \url{https://doi.org/10.1371/journal.pcbi.1012973}






\end{thebibliography}
\end{document}